\documentclass[aps,prd,showpacs,twocolumn,groupedaddress]{revtex4}
\usepackage{epsbox}
\usepackage{graphicx}
\begin{document}

\title{Detailed analysis of the tetraquark potential and flip-flop in SU(3) lattice QCD}

\author{Fumiko~Okiharu}
\affiliation{Department of Physics, Nihon University, 
Kanda-Surugadai 1-8-14, Chiyoda, Tokyo 101-8308, Japan}

\author{Hideo~Suganuma}
\affiliation{Faculty of Science, Tokyo Institute of Technology,
Ohokayama 2-12-1, Meguro, Tokyo 152-8551, Japan}

\author{Toru~T.~Takahashi}
\affiliation{Yukawa Institute for Theoretical Physics, Kyoto University, 
Kitashirakawa-Oiwake, Sakyo, Kyoto 606-8502, Japan}

\date{\today}

\begin{abstract}
We perform the detailed study of the tetraquark (4Q) potential $V_{\rm 4Q}$ 
for various QQ-$\rm \bar{Q}\bar{Q}$ systems 
in SU(3) lattice QCD with $\beta=6.0$ and $16^3 \times 32$ at the quenched level. 
For about 200 different patterns of 4Q systems, 
$V_{\rm 4Q}$ is extracted from the 4Q Wilson loop in 300 gauge configurations, 
with the smearing method to enhance the ground-state component.
We calculate $V_{\rm 4Q}$ for planar, twisted, asymmetric,  
and large-size 4Q configurations, respectively.
Here, the calculation for large-size 4Q configurations 
is done by identifying $16^2 \times 32$ as the spatial size and 
16 as the temporal one, 
and the long-distance confinement force is particularly analyzed in terms of the flux-tube picture. 
When QQ and $\rm \bar{Q}\bar{Q}$ are well separated, 
$V_{\rm 4Q}$ is found to be expressed as 
the sum of the one-gluon-exchange Coulomb term and multi-Y type linear term based on the flux-tube picture.
When the nearest quark and antiquark pair is spatially close, 
the system is described as a ``two-meson" state.
We observe a flux-tube recombination called as ``flip-flop" 
between the connected 4Q state and the ``two-meson" state 
around the level-crossing point.
This leads to infrared screening of the long-range color forces between (anti)quarks belonging to different mesons, 
and results in the absence of the color van der Waals force between two mesons.
\end{abstract}


\maketitle

\section{Introduction}\label{sec1}

The inter-quark force is one of the elementary quantities for the study of 
the multi-quark system in the quark model.
Since the first application of lattice QCD simulations was done 
for the inter-quark potential between a quark and an antiquark using the Wilson loop in 1979 \cite{C7980}, 
the study of the inter-quark force has been one of the important issues in lattice QCD \cite{R97}.
Actually, in hadron physics, the inter-quark force can be regarded as an elementary quantity 
to connect ``the quark world" to ``the hadron world", and plays an important role to hadron properties. 

In addition to the $\rm Q\bar Q$ potential \cite{C7980,R97,APE87,BS93}, recent lattice QCD studies clarify 
the three-quark (3Q) potential \cite{TMNS01,TSNM02,TS0304,SIT04}, which is responsible to the baryon structure. 
In fact, our group recently studied the 3Q potential $V_{\rm 3Q}$ in detail with lattice QCD, 
and clarified that it obeys the Coulomb plus Y-type linear potential \cite{TMNS01,TSNM02,TS0304,SIT04}.
However, no one knows the inter-quark force from QCD in the exotic multi-quark system 
such as tetraquark mesons (QQ-$\bar {\rm Q}\bar{\rm Q}$),  
pentaquark baryons (4Q-$\bar{\rm Q}$), dibaryons (6Q) and so on.

In these years, various candidates of multi-quark hadrons 
have been experimentally observed. 
$\Theta^+$(1540) \cite{LEPS,DIANA,CLAS,SAPHIR}, $\Xi^{--}(1862)$ \cite{H1}
and $\Theta_c(3099)$ \cite{NA49} are considered to be pentaquark (5Q) states \cite{DPP97,Z04}
because of their exotic quantum numbers, although some high-energy experiments report null results\cite{null}.
$X$(3872) \cite{BelleX,CDF,D0,BABARX} and $D_s(2317)$ \cite{BABARDs,BelleDs} 
are expected to be tetraquark (4Q) states \cite{CG03,CP04,T03,PS04,W04,BK04,BG04,ELQ04,S04}
from the consideration of their mass, narrow decay width and decay mode.

These discoveries of multi-quark hadrons are expected to reveal 
new aspects of hadron physics, especially for the inter-quark force such as 
the quark confinement force, the color-magnetic interaction and the diquark correlation \cite{JW03}. 
According to these experimental results, 
it is desired to investigate the inter-quark force 
in the multi-quark system directly based on QCD \cite{GLPM96,PGM99,OST04,STOI04a,STOI04b,SOTI04,SIOT04,OST04p,AK05}, 
which would present the proper Hamiltonian for the quark-model calculation of multi-quarks \cite{WI82,SR03,KMN04}.

As for the 4Q candidates, $X$(3872) was discovered in the process 
$B^+ \rightarrow K^++X(3872) \rightarrow K^++\pi^-\pi^+J/\psi$
at KEK(Belle) in 2003 \cite{BelleX}, and its existence was confirmed by Fermilab (CDF\cite{CDF}, D0\cite{D0}) and SLAC(BABAR)\cite {BABARX}.
$D_s$(2317) was also found in $B$-$\bar{B}$ reaction 
at $\Upsilon(4S)$ resonance at SLAC(BABAR) \cite{BABARDs} 
and consecutively at KEK(Belle) \cite{BelleDs} in 2003. 
As the unusual features of $X$(3872), 
its mass is rather close to the threshold of $D^0(c\bar{u})$ and $\bar{D}^{0*}(u\bar{c})$, 
and its decay width is very narrow as $\Gamma <$ 2.3MeV (90 \% C.L.). 
These facts seem to indicate that $X$(3872) is a 4Q state \cite{CG03,CP04}
or a molecular state of $D^0(c\bar{u})$ and $\bar{D}^{0*}(u\bar{c})$ \cite{T03,PS04,W04,BK04}
rather than an excited-state of a $c\bar c$ system \cite{BG04,ELQ04,S04}.
Also, $D_s$(2317) cannot be regarded as the simple meson of $c\bar{s}$,
but is conjectured to be a 4Q state from the similar reasons on the mass and the narrow decay width.

Also in the light quark sector, the possibility of 4Q hadrons has been pointed out. 
For instance, Jaffe proposed in 1977 that 
the light scalar nonet including $f_0$(980) and $a_0$(980) can be interpreted as 
$\rm QQ\bar{Q}\bar{Q}$ rather than $\rm Q\bar{Q}$ \cite{J77}.
Since then, many studies of the scalar nonet have been done 
in terms of the 4Q state \cite{BFSS99,AJ00}.

As an analytical guiding model for the multi-quark system, 
the flux-tube picture \cite{N74,KS75,CKP83,CNN79,IP8385,P84,O85,LLMRSY86}
has been investigated for the structure and the reaction of hadrons, 
and is supported by recent lattice QCD studies 
\cite{TMNS01,TSNM02,PGM99,OST04,STOI04a,STOI04b,SOTI04,SIOT04,OST04p,IBSS03}.
In this picture, valence quarks are linked 
by the color flux-tube as a quasi-one-dimensional object.
The flux-tube has a large string tension of 
about $\sigma \simeq 0.89$ GeV/fm, 
and therefore its length is to be minimized.
For the multi-quark system, this picture predicts an interesting phenomenon of the ``flip-flop", i.e., 
a recombination of the flux-tube configuration so as to minimize the total length of the flux-tube 
in accordance with the change of the quark location \cite{O85,LLMRSY86}.
This process is important not only for the structure of multi-quark systems 
but also for the reaction process of hadrons.

In this paper, we study the tetraquark (4Q) potential, i.e.,  
the interaction between quarks in the 4Q system 
directly from QCD by using SU(3) lattice QCD at the quenched level, 
and investigate the hypothetical flux-tube picture 
for the multi-quark system and the flip-flop in terms of QCD.
Here, the lattice QCD Monte Carlo simulation is the first-principle calculation of QCD 
and is considered as the only reliable method for nonperturbative QCD at present.
We note that lattice QCD is also a useful method to select out 
the correct picture for nonperturbative QCD in the low-energy region
through the comparison with the lattice results.

The organization of this paper is as follows.
In Sect.II, after a brief review on the lattice studies of static quark potentials, 
we present a theoretical form for the 4Q potential based on the flux-tube picture. 
In Sect.III, we present the formalism for the 4Q Wilson loop and the 4Q potential. 
The lattice QCD results are shown in Sect.IV. 
In Sect.VI, we compare the lattice QCD results with the theoretical form,  
and discuss the flux-tube picture and the flip-flop.  
Sect.VI is devoted for summary and concluding remarks.

\section{Theoretical consideration for the 4Q potential}
\label{theoreticalAnsatz}

\subsection{$\bf Q\bar Q$, 3Q and 5Q potentials}

To begin with, we give a theoretical consideration for the multi-quark potential. 
From a lot of lattice QCD studies \cite{C7980,R97,APE87,BS93,TMNS01,TSNM02}, 
the static Q$\rm \bar{Q}$ potential is known to be well described by 
\begin{eqnarray}
V_{\rm Q\bar{Q}}
&=&-\frac{A_{\rm Q\bar{Q}}}{r}
             +\sigma_{\rm Q\bar{Q}}r+C_{\rm Q\bar{Q}},
\label{QQbar}
\end{eqnarray}
where $r$ denotes the distance between the quark and the antiquark.  
The first term is considered to be 
the Coulomb term due to the one-gluon exchange (OGE) and 
$A_{\rm Q\bar{Q}}$ denotes the Coulomb coefficient.
The second term is the linear confinement term with the string tension, 
$\sigma_{\rm Q\bar{Q}} \simeq$ 0.89 GeV/fm. 

From the recent detailed studies in lattice QCD with 
($\beta$=5.7, $12^3 \times 24$), ($\beta$=5.8, $16^3 \times 32$), ($\beta$=6.0, $16^3 \times 32$), and ($\beta$=6.2, $24^4$) 
\cite{TMNS01,TSNM02,TS0304,SIT04,STOI04a,STOI04b,SOTI04,SIOT04}, 
the 3Q potential is clarified to be 
the sum of the OGE Coulomb term 
and the Y-type linear confinement term as
\begin{eqnarray}
V_{\rm 3Q}=-A_{\rm 3Q}\sum_{i<j}\frac{1}{|{\bf r}_i-{\bf r}_j|}
             +\sigma_{\rm 3Q}L_{\rm min}+C_{\rm 3Q}.
\end{eqnarray}
Here, $L_{\rm min}$ denotes the minimal value of the total flux-tube length, which corresponds to 
the Y-shaped flux-tube linking the three quarks.
In fact, the lattice data of the 3Q potential, $V_{\rm 3Q}^{\rm latt}$, can be well reproduced with 
only three parameters, $A_{\rm 3Q}$, $\sigma_{\rm 3Q}$ and $C_{\rm 3Q}$.
 
To demonstrate the validity of the Y-Ansatz, 
we show in Fig.1 the lattice QCD data of 
the 3Q confinement potential $V_{\rm 3Q}^{\rm conf}$, 
i.e., the 3Q potential subtracted by its Coulomb and constant parts, 
\begin{eqnarray}
V_{\rm 3Q}^{\rm conf} \equiv V_{\rm 3Q}^{\rm latt}-\left\{-A_{\rm 3Q}\sum_{i<j}\frac{1}{|{\bf r}_i-{\bf r}_j|}+C_{\rm 3Q}\right\},
\end{eqnarray}
plotted against Y-shaped flux-tube length $L_{\rm min}$, 
at $\beta$=5.8, 6.0 and 6.2 in the lattice unit.
For each $\beta$, clear linear correspondence is found between 3Q confinement potential 
$V_{\rm 3Q}^{\rm conf}$ and $L_{\rm min}$ as $V_{\rm 3Q}^{\rm conf} \simeq \sigma_{\rm 3Q} L_{\rm min}$,
which indicates the Y-Ansatz for the 3Q potential \cite{STOI04b,SOTI04,SIOT04}.

\begin{figure}[h]
\begin{center}
\includegraphics[height=7cm]{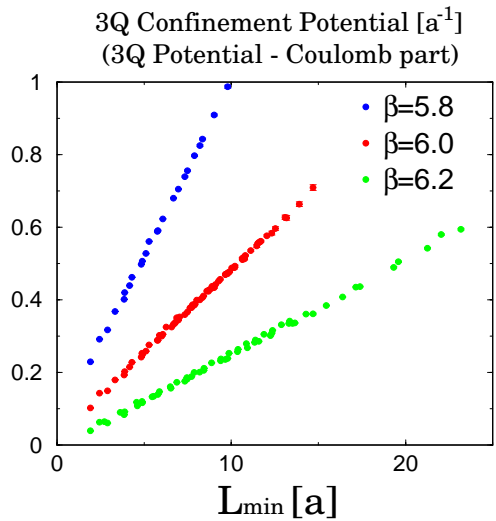} 
\caption{
The lattice QCD result for the 3Q confinement potential $V_{\rm 3Q}^{\rm conf}$, 
i.e., the 3Q potential subtracted by its Coulomb and constant parts, 
plotted against Y-shaped flux-tube length $L_{\rm min}$ 
at $\beta$=5.8, 6.0 and 6.2 in the lattice unit.
The clear linear correspondence between 3Q confinement potential 
$V_{\rm 3Q}^{\rm conf}$ and $L_{\rm min}$
indicates the Y-Ansatz for the 3Q potential.
}
\end{center}
\label{Fig1}
\end{figure}

This lattice QCD result strongly supports the flux-tube picture for baryons, and 
the Y-type flux-tube formation is actually observed in lattice QCD 
through the direct measurement of the action density 
in the presence of spatially-fixed three quarks 
\cite{STOI04a,STOI04b,SOTI04,SIOT04,IBSS03}.
The Y-Ansatz for the 3Q system is also supported by recent further lattice QCD studies \cite{BIMPS04,BS04} 
and analytical studies \cite{KS03,C04}.

As for the relation between $V_{\rm Q\bar Q}$ and $V_{\rm 3Q}$, 
we have found the OGE result $A_{\rm 3Q} \simeq \frac12 A_{\rm Q\bar Q}$ and 
the universality of the string tension $\sigma_{\rm 3Q} \simeq \sigma_{\rm Q\bar Q}$, 
which also supports the flux-tube picture 
\cite{N74,KS75,CKP83,CNN79,IP8385,P84,O85,LLMRSY86}
and the strong-coupling expansion scheme \cite{KS75,CKP83}.

Very recently, the 5Q potential is also studied in lattice QCD
\cite{OST04,STOI04a,STOI04b,SOTI04,SIOT04,OST04p,AK05}. 
It is well described 
by the OGE Coulomb plus multi-Y type linear potential \cite{OST04,STOI04a,STOI04b,SOTI04,SIOT04,OST04p}.
With the minimal length $L_{\rm min}$ of the flux-tube, 
the 5Q potential can be well described as 
\begin{eqnarray}
&&V_{\rm 5Q}
=\frac{g^2}{4\pi} \sum_{i<j} \frac{T^a_i T^a_j}{|{\bf r}_i-{\bf r}_j|}+\sigma_{\rm 5Q} L_{\rm min}+C_{\rm 5Q} \nonumber \\
&=&-A_{\rm 5Q}\{ ( \frac1{r_{12}}  + \frac1{r_{34}}) 
+\frac12(\frac1{r_{15}} +\frac1{r_{25}} +\frac1{r_{35}} +\frac1{r_{45}}) \nonumber \\
&+&\frac14(\frac1{r_{13}} +\frac1{r_{14}} +\frac1{r_{23}} +\frac1{r_{24}}) \}
+\sigma_{\rm 5Q} L_{\rm min}+C_{\rm 5Q},
\label{V5Qth}
\end{eqnarray}
with $(A_{\rm 5Q}, \sigma_{\rm 5Q})$ fixed to be $(A_{\rm 3Q}, \sigma_{\rm 3Q})$ following 
the OGE result and the universality of the string tension. 
This lattice result also supports the flux-tube picture.

\subsection{Theoretical Ans\"atze for the 4Q potential}

Now, we investigate the theoretical form of the tetraquark (4Q) potential $V_{\rm 4Q}$ 
with respect to the flux-tube picture, which seems workable for $\rm Q\bar Q$ mesons and baryons.
For the argument of the low-lying 4Q states, 
we consider the 4Q state of ((QQ)$_{\bar{3}}$-($\rm \bar{Q}\bar{Q}$)$_3$)${_1}$ 
as shown in Fig.2. 
Here, (QQ)$_{\bar{3}}$ denotes that two quarks form 
the $\rm \bar{\bf 3}$ representation of the color SU(3). 
The meaning of ($\rm \bar{Q}\bar{Q}$)$_3$ is similar.
By combining (QQ)$_{\bar{\bf 3}}$ with ($\rm \bar{Q}\bar{Q}$)$_3$, 
the 4Q system can be constructed as a color-singlet state.
We note that another possible 4Q system of ((QQ)$_6$-$(\bar{\rm Q}\bar{\rm Q}$)$_{\bar{6}})_1$ is 
expected to be a highly-excited state, since 
the attractive (repulsive) force acts between quarks, 
when the QQ cluster belongs to $\bar{\bf 3}$ (${\bf 6}$) representation
in a perturbative sense, which leads to the $\bar{\bf 3}$ diquark picture \cite{JW03}.

\begin{figure}[h]
\begin{center}
\includegraphics[width=5cm]{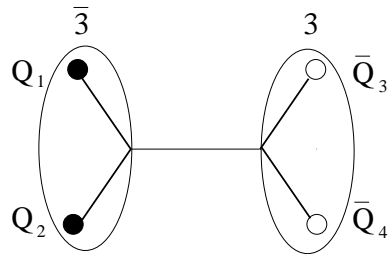}
\end{center}
\caption
{The tetraquark (QQ-$\rm \bar{Q}\bar{Q}$) state.
The QQ ($\rm \bar{Q}\bar{Q}$) cluster
belongs to $\bar{\bf 3}$ ({\bf 3}) representation 
of the color SU(3). 
}
\label{Fig2}
\end{figure}

In the flux-tube picture, the flux-tube is formed so as to minimize the total flux-tube length of the system 
for the low-lying state.
For the low-lying 4Q system, there are two candidates for the flux-tube configuration according to the 4Q location.
One is the connected flux-tube system where
all quarks and antiquarks are connected with the single flux-tube 
as shown in Fig.3. 
The other is the disconnected flux-tubes corresponding to a ``two-meson" state 
as shown in Fig.4.
Note that  the 4Q state of ((QQ)$_{\bar{3}}$-($\rm \bar{Q}\bar{Q}$)$_3$)${_1}$ generally includes 
such a ``two-meson" state of $(\rm Q \bar Q)_1 (\rm Q \bar Q)_1$ as shown in Appendix A.
For each case, we consider below the theoretical form of the tetraquark potential $V_{\rm 4Q}$.

\begin{figure}[h]
\begin{center}
\includegraphics[height=4cm]{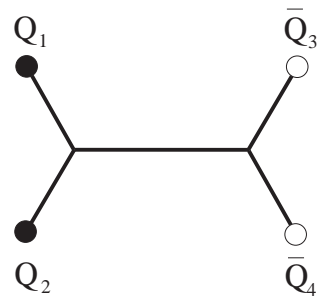}
\end{center}
\caption
{The connected tetraquark system.
All quarks and antiquarks are connected with the single flux-tube, which is double-Y-shaped. 
}
\label{Fig3}
\end{figure}
\begin{figure}[h]
\begin{center}
\includegraphics[height=4cm]{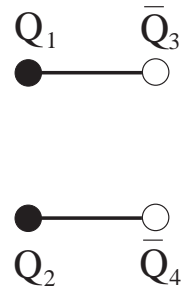}
\end{center}
\caption
{The disconnected tetraquark system, which corresponds to a ``two-meson" state.  
}
\label{Fig4}
\end{figure}

\subsubsection{OGE plus multi-Y Ansatz for the connected 4Q system.}

For the connected 4Q system, we propose the ``OGE plus multi-Y Ansatz" as a theoretical form of $V_{\rm 4Q}$
from the viewpoint of the flux-tube picture. 
This type of the flux-tube configuration is plausible 
when the distance between QQ and $\rm \bar Q\bar Q$ is enough long compared with the size of these clusters.
For such a system, all quarks and antiquarks are linked by the connected double-Y-shaped flux-tube as shown in Fig.3, 
and the 4Q potential $V_{\rm 4Q}$ is described by 
the OGE Coulomb plus multi-Y linear potential $V_{\rm c4Q}$ as
\begin{eqnarray}
V_{\rm 4Q}
&=&\frac{g^2}{4\pi}\sum_{i<j}\frac{T^a_i T^a_j}{|r_i-r_j|}
     +\sigma_{\rm 4Q}L_{\rm min}+C_{\rm 4Q} \nonumber \\
&=&-A_{\rm 4Q}\{(\frac{1}{r_{12}}+\frac{1}{r_{34}})
    +\frac{1}{2}(\frac{1}{r_{13}}+\frac{1}{r_{14}} \nonumber \\
& &             +\frac{1}{r_{23}}+\frac{1}{r_{24}}) \}
   +\sigma_{\rm 4Q}L_{\rm min}+C_{\rm 4Q} \equiv V_{\rm c4Q}
\label{Vc4Q}
\end{eqnarray}
with $r_{ij} \equiv |{\bf r}_i-{\bf r}_j|$ and 
$L_{\rm min}$ being the minimal value of the total flux-tube length.
Here, ${\bf r}_i$ denotes the location of $i$th (anti)quark in Fig.3.

The first term describes the Coulomb term due to the OGE process.  
Note that there appear two kinds of Coulomb coefficients ($A_{\rm 4Q}$, $\frac12 A_{\rm 4Q}$) in the 4Q system, 
while only one Coulomb coefficient, $A_{\rm Q\bar {\rm Q}}$ or $A_{\rm 3Q}$, appears in the Q$\bar{\rm Q}$ or the 3Q system. 
In this definition, the Coulomb coefficient $A_{\rm 4Q}$ is expected to satisfy 
$A_{\rm 4Q} \simeq \frac{1}{2}A_{\rm Q\bar{Q}}$ as the OGE results.
The brief derivation of the OGE Coulomb terms is shown in Appendix A. 

The second term is the linear confinement term with the string tension $\sigma_{\rm 4Q}$, 
which is expected to satisfy $\sigma_{\rm 4Q} \simeq \sigma_{\rm Q \bar Q} \simeq$0.89 GeV/fm
as the universality of the string tension.
Similar to the 3Q and the 5Q systems, the Y-type junction appears in this case, 
and $L_{\rm min}$ is expressed by the length of the double-Y-shaped flux-tube as shown in Fig.3.

In the extreme case, e.g., $r_{12}=r_{34} \gg r_{13}=r_{24}$, the lowest connected 4Q system takes an X-shaped flux-tube,  
although the energy of such system is larger than that of the two-meson state 
in most cases.

\subsubsection{The two-meson Ansatz for the disconnected 4Q system}

For the disconnected 4Q system corresponding to the ``two-meson" state as shown in Fig.4, 
we adopt the ``two-meson Ansatz" for $V_{\rm 4Q}$.
Such a flux-tube configuration is plausible 
when the nearest quark and antiquark pair is spatially close and the system can be 
regarded as the ``two-meson state" rather than an inseparable 4Q state.
For such a system as shown in Fig.4, the total potential $V_{\rm 4Q}$ 
for the 4Q system would be approximated to be the sum of two Q$\rm \bar{Q}$ potentials in Eq.(\ref{QQbar}) as 
\begin{eqnarray}
V_{\rm 4Q}
&=&V_{\rm Q\bar{Q}}(r_{13}) + V_{\rm Q\bar{Q}}(r_{24}) \nonumber \\
&=&-A_{\rm Q\bar{Q}}(\frac{1}{r_{13}}+\frac{1}{r_{24}})
           +\sigma_{\rm Q\bar{Q}}(r_{13}+r_{24}) +2C_{\rm Q\bar{Q}} \nonumber \\
&\equiv& V_{\rm 2Q\bar Q},
\label{V2QQbar}
\end{eqnarray} 
assuming that the inter-meson force is subdominant.

\subsection{The 4Q potential form  and the flip-flop} 

For the lowest 4Q system, the 4Q potential $V_{\rm 4Q}$ would be expressed with 
lower energy of the connected 4Q system or the two-meson system, 
\begin{eqnarray}
V_{\rm 4Q}={\rm min}(V_{\rm c4Q}, V_{2\rm Q\bar Q}).
\label{4Qth}
\end{eqnarray}
 
As a physical consequence of Eq.(\ref{4Qth}) based on the flux-tube picture, 
there can occur a physical transition between 
the connected 4Q state and the two-meson state 
according to the change of the 4Q location.
This phenomenon occurs through the recombination of the flux-tube and is called as the ``flip-flop".
(A popular usage of the ``flip-flop" may be for the simple flux-tube recombination between two-meson states.
We here use this term as the general flux-tube recombination.)

The flip-flop is important for the properties of 4Q states especially 
for the decay process into two mesons. 
In addition, the flux-tube recombination between two-meson states can be realized 
through the two successive flip-flops between the two-meson state and the connected 4Q state.
Therefore, this type of the flip-flop is important 
also for the reaction mechanism between two mesons.

\section{Formalism for the 4Q potential in lattice QCD}
\label{formalism}

In this section, we present the formalism to extract the static 4Q potential.
Similar to the derivation of the Q$\rm\bar{Q}$ potential 
from the Wilson loop, 
the 4Q potential $V_{\rm 4Q}$ can be derived 
from the gauge-invariant 4Q Wilson loop $W_{\rm 4Q}$ 
as shown in Fig.5 \cite{STOI04b,OST04p,AK05}.

\begin{figure}[h]
\begin{center}
\includegraphics[width=7cm]{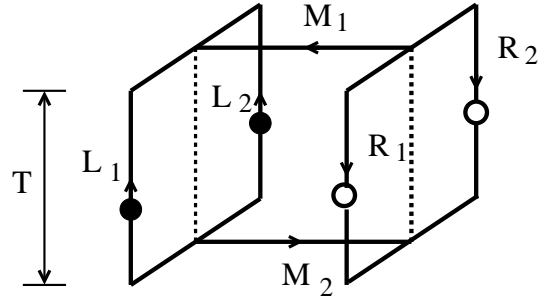}
\end{center}
\caption
{The tetraquark (4Q) Wilson loop 
for the calculation of the 4Q potential $V_{\rm 4Q}$.
The contours $M_i$ ($i$=1,2) are line-like, 
and $L_i,R_i$ ($i$=1,2) are staple-like.
The 4Q gauge-invariant state is generated at $t=0$ 
and is annihilated at $t=T$.
The four quarks ($\rm QQ \bar Q \bar Q$) are spatially fixed 
in $\rm \bf R^3$ for $0<t<T$.}
\label{Fig5}
\end{figure}

The 4Q Wilson loop is defined by 
\begin{eqnarray}
W_{\rm 4Q} \equiv 
\frac{1}{3} {\rm tr} 
(\tilde{M}_1 \tilde{L}_{12} \tilde{M}_2 \tilde{R}_{12}).
\end{eqnarray}
$\tilde{M}_i$, $\tilde{L_i}$,$\tilde{R_i}$ ($i$=1,2) are given by 
\begin{eqnarray}
\tilde{M}_i,\tilde{L_i}, \tilde{R_i}
\equiv 
P \exp{\{ig \int_{M_i,L_i,R_i}dx^\mu A_\mu (x)\}}\in \rm{SU(3)_c}.
\end{eqnarray}
Here, $\tilde{L}_{12}$ and $\tilde{R}_{12}$ are defined by
\begin{eqnarray}
\tilde{L}_{12}^{a'a} 
\equiv \frac{1}{2}\epsilon^{abc}\epsilon^{a'b'c'}
L_1^{bb'}L_2^{cc'}, \\
\tilde{R}_{12}^{a'a} 
\equiv \frac{1}{2}\epsilon^{abc}\epsilon^{a'b'c'}
R_1^{bb'}R_2^{cc'}.
\end{eqnarray}
The ground-state 4Q potential $V_{\rm 4Q}$ is extracted as 
\begin{eqnarray}
V_{\rm 4Q}=-\lim_{T \rightarrow \infty} 
\frac{1}{T} \ln \langle W_{\rm 4Q} \rangle.
\end{eqnarray}

In general, the 4Q Wilson loop $\langle W_{\rm 4Q} \rangle$ 
contains excited-state contributions, and is expressed  as
\begin{eqnarray}
\langle W_{4Q} \rangle = \sum_{n=0}^\infty C_n e^{-V_n T},
\end{eqnarray}
where $V_0$ denotes the ground-state 4Q potential $V_{\rm 4Q}$ and 
$V_{n}$ $(n=1,2,3..)$ the $n$th excited-state potential. 
In principle, $V_{\rm 4Q}$ can be obtained from the behavior of 
$\langle W_{\rm 4Q} \rangle$ at the large $T$ region where 
the ground-state contribution becomes dominant.
In the practical simulation, however, 
the accurate measurement of $V_{\rm 4Q}$ is not easy for large $T$, 
since $\langle W_{\rm 4Q} \rangle$ decreases exponentially with $T$.

To extract the ground-state potential $V_{\rm 4Q}$ in lattice QCD, 
we adopt the gauge-covariant smearing method \cite{R97,APE87,BS93,TMNS01,TSNM02,TS0304} to enhance 
the ground-state component of the 4Q state in the 4Q Wilson loop.
The smearing is known to be a powerful method for the accurate 
measurement of the Q-$\bar {\rm Q}$ \cite{R97,APE87,BS93} 
and the 3Q potentials \cite{TMNS01,TSNM02,TS0304}, 
and is expressed as the iterative replacement of 
the spatial link variables $U_i(s)$ ($i$=1,2,3) 
by the obscured link variables $\bar U_i(s)\in {\rm SU}(3)_c$ 
which maximizes
${\rm Re} \,{\rm tr} \,\,\{\bar U_i^{\dagger}(s) V_i(s)\}$ with
\begin{eqnarray}
V_i(s)\equiv 
\alpha U_i(s)+\sum_{j \ne i} \sum_{\pm} 
\{ U_{\pm j}(s)U_i(s\pm \hat j)U_{\pm j}^\dagger (s+\hat i) \}~~~ 
\label{smearing}
\end{eqnarray}
with the simplified notation of $U_{-j}\equiv U^\dagger_{j}(s-\hat j)$. 
We here adopt the smearing parameter $\alpha=2.3$ 
and the iteration number $N_{\rm smr}=30$, which enhance 
the ground-state component in the 4Q Wilson loop at $\beta$=6.0 in most cases. 
(See the next section.)

\section{Lattice QCD results for the 4Q potential}
\label{results}

The lattice QCD simulations are performed with the standard plaquette action 
at $\beta=6.0$ on the $16^3 \times 32$ lattice at the quenched level. 
The lattice spacing $a$ is estimated as $a \simeq$ 0.104fm, which leads to 
the string tension $\sigma_{\rm Q\bar Q} \simeq (427{\rm MeV})^2$ 
in the Q$\rm \bar Q$ potential, using 
the numerical relation $\sigma_{\rm Q\bar Q} \simeq 0.0506a^{-2}$ obtained 
from the fitting analysis on the on-axis data of the Q$\bar {\rm Q}$ potential 
in lattice QCD at $\beta=6.0$ \cite{TSNM02,SIT04}.
The gauge configurations are taken every 500 sweeps after 
5000 sweeps using the pseudo-heat-bath algorism.
We use 300 configurations for the 4Q potential simulation.
For the estimation of the statistical error of the lattice data, 
we adopt the jack-knife error estimate.

On the $16^3 \times 32$ lattice, 
we investigate the typical configuration of 4Q systems 
as shown in Figs.6 and 7. 
In Fig.6, the 4Q system has a planar structure.
In Fig.7, the 4Q system has a twisted (three-dimension) structure.
In particular, we analyze in detail the symmetric planar and twisted 
4Q configurations with $d_1=d_2=d_3=d_4\equiv d$, 
although more general asymmetric 4Q configurations with various $(d_1,d_2,d_3,d_4)$ 
are also investigated.

For the 4Q configurations with $h \le 8$, 
we identify $16^3$ as the spatial size and 32 as the temporal one.
On the other hand, the calculation for 
the large-size 4Q configurations with $h > 8$ is performed 
by identifying $16^2 \times 32$ as the spatial size and 
16 as the temporal one.
In both cases, 
we use corresponding translational and rotational symmetries on lattices 
for the calculation of $\langle W_{\rm 4Q}\rangle$.

\begin{figure}[h]
\begin{center}
\includegraphics[width=5cm]{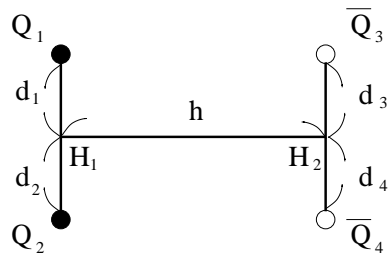}
\end{center}
\caption{A planar configuration of the tetraquark system.
$\rm Q_1 Q_2$ is parallel to $\rm \bar{Q}_3 \bar{Q}_4$, and
$\rm H_1 H_2$ is perpendicular to 
$\rm Q_1 Q_2$ and $\rm \bar{Q}_3 \bar{Q}_4$.
We call the cases with $d_1=d_2=d_3=d_4\equiv d$ 
as ``symmetric planar 4Q configurations".}
\label{Fig6}
\end{figure}
\begin{figure}[h]
\begin{center}
\includegraphics[width=5cm]{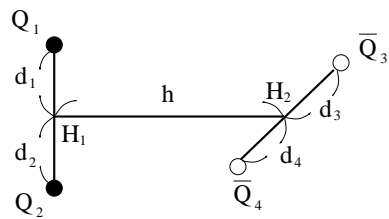}
\end{center}
\caption{A twisted configuration of the tetraquark system.
$\rm Q_1 Q_2$ is perpendicular to $\rm \bar{Q}_3 \bar{Q}_4$, and 
$\rm H_1 H_2$ is perpendicular to 
$\rm Q_1 Q_2$ and $\rm \bar{Q}_3 \bar{Q}_4$.
We call the cases with $d_1=d_2=d_3=d_4\equiv d$ 
as ``symmetric twisted 4Q configurations".}
\label{Fig7}
\end{figure}

For these types of 4Q configurations, 
we construct the 4Q Wilson loop $W_{\rm 4Q}$ with 
the junctions locating at ${\rm H}_1$ and ${\rm H}_2$, 
and calculate the 4Q potential $V_{\rm 4Q}$ from 
$\langle W_{\rm 4Q}\rangle$ using the smearing method.

For the suitable choice of the smearing parameter $\alpha$ 
and the iteration number $N_{\rm smr}$ in Eq.(\ref{smearing}), 
we perform some numerical tests with various values of $\alpha$ and $N_{\rm smr}$, 
and finally adopt $\alpha=2.3$ and $N_{\rm smr}=30$, which are found to enhance 
the ground-state component in the 4Q Wilson loop at $\beta$=6.0 in most cases.
For the demonstration, we show in Fig.8 a typical example of 
the ground-state-overlap quantity 
\begin{eqnarray}
C_0 \equiv \langle W_{\rm 4Q}(T)\rangle^{T+1}/\langle W_{\rm 4Q}(T+1)\rangle^{T}
\end{eqnarray}
plotted against the iteration number $N_{\rm smr}$ at $\alpha=2.3$
for the symmetric planar 4Q configuration with $(d,h)=(1,5)$. 
Here, the ground-state-overlap quantity $C$ indicates the magnitude of 
the ground-state component \cite{TMNS01,TSNM02}, and 
is found to take a large value close to unity around $N_{\rm smr}=30$. 
(Note here that the enhancement of the ground-state component 
is the aim of the smearing, and hence any approximate optimization is 
applicable as long as the ground-state overlap is enough large.)

\begin{figure}[h]
\begin{center}
\includegraphics[height=5.5cm]{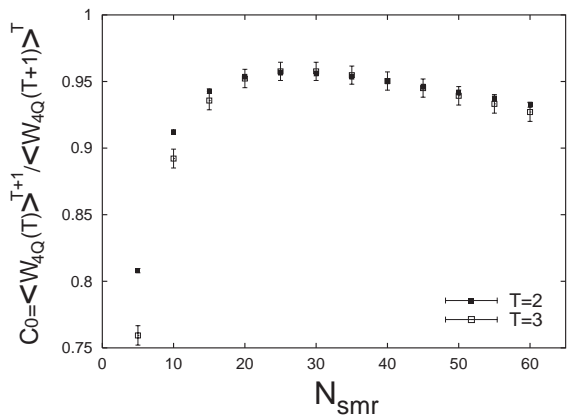}
\end{center}
\caption{A typical example of the ground-state-overlap quantity 
$C_0 \equiv \langle W_{\rm 4Q} (T) \rangle^{T+1}/\langle W_{\rm 4Q}(T+1) \rangle^{T}$
plotted against the iteration number $N_{\rm smr}$ at $\alpha=2.3$
for the symmetric planar 4Q configuration with $(d,h)=(1,5)$. 
$C_0$ takes a large value close to unity around $N_{\rm smr}=30$. 
}
\label{Fig8}
\end{figure}

Due to the smearing, the ground-state component is largely enhanced in most cases, 
and therefore the 4Q Wilson loop $\langle W_{\rm 4Q} \rangle$ 
composed with the smeared link-variable exhibits 
a single-exponential behavior as 
$\langle W_{\rm 4Q} \rangle \simeq e^{-V_{\rm 4Q}T}$ 
even for a small value of $T$. 
Then, for each 4Q configuration, we extract $V_{\rm 4Q}$
from the least squares fit with the single-exponential form
\begin{equation}
\langle W_{\rm 4Q}\rangle =\bar{C}e^{-V_{\rm 4Q}T}
\label{expfit}
\end{equation}
in the range of $T_{\rm min}\leq T\leq T_{\rm max}$ listed in Table~I-VI.
The prefactor $\bar C$ physically means the ground-state overlap, 
and $\bar C\simeq 1$ corresponds to the realization of the quasi-ground-state.
Here, we choose the fit range of $T$ such that the stability of the
``effective mass'' 
\begin{eqnarray}
V(T)\equiv \ln\{\langle W_{\rm 4Q}(T) \rangle /
\langle W_{\rm 4Q}(T+1)\rangle\}
\end{eqnarray}
is observed in the range of $T_{\rm min}\leq T\leq T_{\rm max}-1$.

To see how excited-state contamination is removed in this calculation, 
we show in Fig.9 several effective-mass plots, $V(T)$ v.s. $T$,  
for planar and twisted 4Q configurations 
at small, intermediate and large distances, respectively.
Owing to the smearing, the effective mass $V(T)$ seems to be stable even for small $T$.
To show the quality of the single-exponential fit for $\langle W_{\rm 4Q}\rangle$ as in Eq.(\ref{expfit}), 
we list the chi-square per degree of freedom, 
$\chi^2/N_{\rm DF}$, for each fit in Table I-VI. 
For most cases, $\chi^2/N_{\rm DF}$ takes a small value less than unity, and 
the fitting seems to be plausible.
(We note that the errors listed in Table I-VI are the statistical ones, and 
there are some systematical errors in the lattice QCD calculation.
For instance, when $T_{\rm min}=2$ is adopted for large 4Q systems as $(d,h)$=(3,16), 
the systematical error originating from the fit-range choice seems 
to be several times larger than the statistical error.)

\begin{figure}[h]
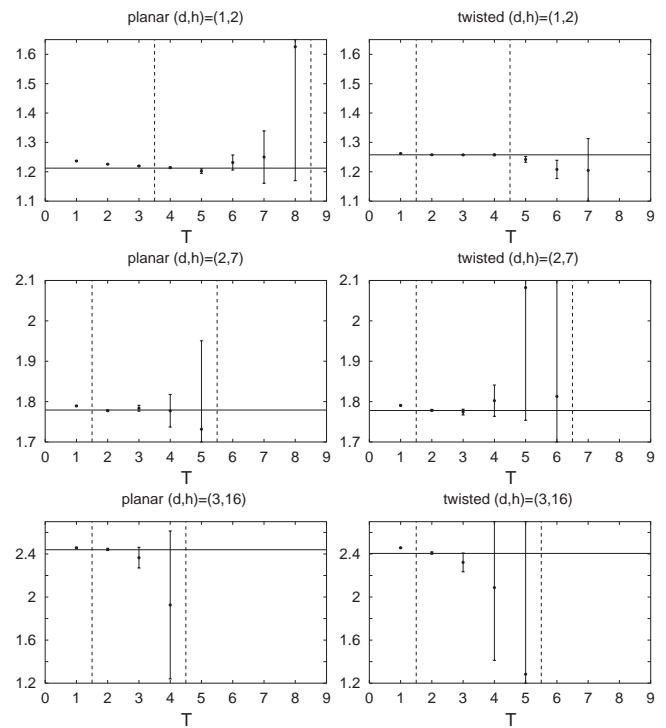

\begin{center}
\includegraphics[width=4.2cm]{Fig9a.eps}
\includegraphics[width=4.2cm]{Fig9b.eps}
\includegraphics[width=4.2cm]{Fig9c.eps}
\includegraphics[width=4.2cm]{Fig9d.eps}
\includegraphics[width=4.2cm]{Fig9e.eps}
\includegraphics[width=4.2cm]{Fig9f.eps}
\end{center}
\caption{
The effective-mass plots, $V(T)$ v.s. $T$, for several 4Q configurations 
at small, intermediate and large distances:
the symmetric planar and twisted 4Q configurations with $(d,h)$=(1,2), $(d,h)$=(2,7), and $(d,h)$=(3,16).
The fit range of $T_{\rm min} \le T \le T_{\rm max}-1$, which is regarded as the plateau region, 
is indicated by the vertical dashed lines.
The solid horizontal line denotes the final fit of $V_{\rm 4Q}$ from the least square fit with 
the single-exponential form. 
}
\label{Fig9}
\end{figure}

In this way, 
we calculate the tetraquark potential $V_{\rm 4Q}$ for various 4Q system, {\it i.e.,}  
planar, twisted, asymmetric, and large-size 4Q configurations, respectively.
We summarize in Table~I-VI  
the lattice QCD results for $V_{\rm 4Q}$ together with 
the ground-state overlap $\bar C$, the fit range of $T$, $\chi^2/N_{\rm DF}$, 
the minimal flux-tube length $L_{\rm min}^{\rm c4Q}$ for the connected 4Q configuration, 
and the theoretical Ans\"atze $V_{\rm c4Q}^{\rm th}$ and $V_{\rm 2Q\bar Q}^{\rm th}$ 
presented in Sect.II.
\begin{enumerate}
\item
Table~I and II show $V_{\rm 4Q}$ for the symmetric planar 4Q configurations
as shown in Fig.6 with $d_1=d_2=d_3=d_4\equiv d$. 
$V_{\rm 4Q}$ is shown in terms of $d$ and $h$.
\item
Table~III and IV show $V_{\rm 4Q}$ for the symmetric twisted 4Q configurations
as shown in Fig.7 with $d_1=d_2=d_3=d_4\equiv d$.
$V_{\rm 4Q}$ is shown in terms of $d$ and $h$.
\item
Table~V shows $V_{\rm 4Q}$ for the asymmetric planar 4Q configurations
as shown in Fig.6 with various $(d_1,d_2,d_3,d_4)$ for $h$=8.
\item 
Table~VI shows $V_{\rm 4Q}$ for the asymmetric twisted 4Q configurations
as shown in Fig.7 with various $(d_1,d_2,d_3,d_4)$ for $h$=8.
\end{enumerate} 
Thus, we obtain the tetraquark potential $V_{\rm 4Q}$ 
for about 200 different patterns of 4Q systems.

\begin{table*}[h]
\caption{
A part of lattice QCD results of the 4Q potential $V_{\rm 4Q}$ for 
the planar 4Q configuration as shown in Fig.6 with $d_1=d_2=d_3=d_4\equiv d$. 
The symmetric planar 4Q systems are labeled by $(d,h)$.
We list also the ground-state overlap $\bar C$, the fit range of $T$, $\chi^2/N_{\rm DF}$, 
the minimal flux-tube length $L_{\rm min}^{\rm c4Q}$ for the connected 4Q configuration, 
and the theoretical Ans\"atze.
$V_{\rm c4Q}^{\rm th}$ denotes the OGE Coulomb plus multi-Y Ansatz (\ref {Vc4Q}) 
with $(A_{\rm 4Q},\sigma_{\rm 4Q})$ fixed to be $(A_{\rm 3Q},\sigma_{\rm 3Q})$ in $V_{\rm 3Q}$ in Ref.\cite{TSNM02}. 
$V_{\rm 2Q\bar{Q}}^{\rm th}$ denotes the two-meson Ansatz (\ref{V2QQbar}) with $V_{\rm Q\bar Q}$ in Ref.\cite{TSNM02}. 
All the quantities are measured in the lattice unit at $\beta=6.0$, i.e., $a \simeq$ 0.104fm.
}
\label{Table1}
\begin{ruledtabular}
\begin{tabular}{cccccccc} 
($d,h$) & $V_{\rm 4Q}$     & $\bar C$ 
& $T_{\rm min}-T_{\rm max}$ 
& $\chi^2/N_{\rm DF}$
& $L_{\rm min}^{\rm c4Q}$
& $V_{\rm c4Q}^{\rm th}$  
& $V_{\rm 2Q\bar{Q}}^{\rm th}$ 
\\ \hline
(1,1) & 0.8590(45) & 0.4970(135) & 6-11 & 0.140 &  4.47 & 1.1293 & 0.8224\\
(1,2) & 1.2124(31) & 0.9049(111) & 4-9 & 0.710 &  5.46 & 1.2560 & 1.2004\\
(1,3) & 1.3218(17) & 0.9572(48) & 3-9 & 0.195 &  6.46 & 1.3352 & 1.3939\\
(1,4) & 1.3938(11) & 0.9627(19) & 2-8 & 0.755 &  7.46 & 1.3999 & 1.5412\\
(1,5) & 1.4531(13) & 0.9556(22) & 2-8 & 0.633 &  8.46 & 1.4579 & 1.6701\\
(1,6) & 1.5051(33) & 0.9354(95) & 3-9 & 0.072 &  9.46 & 1.5123 & 1.7897\\
(1,7) & 1.5644(19) & 0.9434(32) & 2-7 & 0.336 & 10.46 & 1.5643 & 1.9041\\
(1,8) & 1.6184(20) & 0.9375(36) & 2-8 & 0.124 & 11.46 & 1.6150 & 2.0152\\
(1,9) & 1.6706(22) & 0.9295(35) & 2-4 & 0.119 & 12.46 & 1.6646 & 2.1241\\
(1,10) & 1.7269(28) & 0.9286(48) & 2-7 & 0.184 & 13.46 & 1.7136 & 2.2314\\
(1,11) & 1.7774(30) & 0.9177(48) & 2-4 & 0.001 & 14.46 & 1.7620 & 2.3377\\
(1,12) & 1.8234(37) & 0.8983(62) & 2-7 & 1.094 & 15.46 & 1.8100 & 2.4431\\
(1,13) & 1.8797(44) & 0.8979(74) & 2-7 & 0.351 & 16.46 & 1.8577 & 2.5478\\
(1,14) & 1.9285(49) & 0.8838(78) & 2-6 & 0.899 & 17.46 & 1.9052 & 2.6521\\
(1,15) & 1.9787(55) & 0.8722(91) & 2-7 & 0.361 & 18.46 & 1.9525 & 2.7559\\
(1,16) & 2.0365(62) & 0.8741(101) & 2-6 & 0.702 & 19.46 & 1.9996 & 2.8594\\
(2,1) & 0.8263(46) & 0.3324(93) & 6-13 & 0.624 &  8.25 & 1.3992 & 0.8224\\
(2,2) & 1.2633(183) & 0.5646(514) & 5-9 & 0.727 &  8.94&1.5022 & 1.2004\\
(2,3) & 1.4592(111) & 0.7457(328) & 4-7 & 0.119 &  9.93&1.5734 & 1.3939\\
(2,4) & 1.5950(17) & 0.9223(29) & 2-7 & 0.643 & 10.93 & 1.6340 & 1.5412\\
(2,5) & 1.6619(21) & 0.9286(36) & 2-6 & 0.779 & 11.93 & 1.6896 & 1.6701\\
(2,6) & 1.7215(23) & 0.9285(41) & 2-7 & 0.653 & 12.93 & 1.7426 & 1.7897\\
(2,7) & 1.7791(29) & 0.9283(51) & 2-6 & 0.089 & 13.93 & 1.7938 & 1.9041\\
(2,8) & 1.8279(31) & 0.9121(54) & 2-6 & 0.631 & 14.93 & 1.8439 & 2.0152\\
(2,9) & 1.8827(34) & 0.9101(57) & 2-7 & 0.111 & 15.93 & 1.8932 & 2.1241\\
(2,10) & 1.9366(38) & 0.9049(64) & 2-6 & 0.108 & 16.93 & 1.9419 & 2.2314\\
(2,11) & 1.9917(45) & 0.9026(73) & 2-5 & 0.785 & 17.93 & 1.9902 & 2.3377\\
(2,12) & 2.0419(49) & 0.8911(82) & 2-5 & 0.215 & 18.93 & 2.0381 & 2.4431\\
(2,13) & 2.0954(55) & 0.8856(90) & 2-6 & 0.445 & 19.93 & 2.0857 & 2.5478\\
(2,14) & 2.1387(61) & 0.8619(101) & 2-5 & 1.272 & 20.93 & 2.1331 & 2.6521\\
(2,15) & 2.1990(67) & 0.8683(111) & 2-5 & 1.103 & 21.93 & 2.1804 & 2.7559\\
(2,16) & 2.2482(78) & 0.8567(122) & 2-5 & 0.687 & 22.93 & 2.2274 & 2.8594\\
\end{tabular}
\end{ruledtabular}
\end{table*}

\begin{table*}[t]
\caption{
A part of lattice QCD results of the 4Q potential $V_{\rm 4Q}$ 
for the symmetric planar 4Q configuration labeled by $(d,h)$. 
The notations are the same as in Table~I.}
\label{Table2}
\begin{ruledtabular}
\begin{tabular}{cccccccc} 
($d,h$) & $V_{\rm 4Q}$     & $\bar C$ & $T_{\rm min}-T_{\rm max}$ 
& $\chi^2/N_{\rm DF}$
& $L_{\rm min}^{\rm c4Q}$ 
& $V_{\rm c4Q}^{\rm th}$ 
& $V_{\rm 2Q\bar{Q}}^{\rm th}$\\ \hline
(3,1) & 0.8281(20) & 0.3242(33) & 5-12 & 0.502 & 12.17 & 1.6129 & 0.8224\\
(3,2) & 1.2143(228) & 0.3390(387) & 5-9 & 0.237 & 12.65&1.7043 & 1.2004\\
(3,3) & 1.5031(168) & 0.5540(370) & 4-7 & 2.112 & 13.42&1.7636 & 1.3939\\
(3,4) & 1.6992(68) & 0.7751(158) & 3-7 & 0.654 & 14.39 & 1.8213 & 1.5412\\
(3,5) & 1.8185(31) & 0.8797(53) & 2-6 & 0.446 & 15.39 & 1.8756 & 1.6701\\
(3,6) & 1.8896(35) & 0.8922(60) & 2-6 & 0.299 & 16.39 & 1.9275 & 1.7897\\
(3,7) & 1.9459(37) & 0.8849(63) & 2-4 & 1.708 & 17.39 & 1.9780 & 1.9041\\
(3,8) & 2.0016(45) & 0.8808(76) & 2-5 & 0.933 & 18.39 & 2.0276 & 2.0152\\
(3,9) & 2.0544(45) & 0.8739(73) & 2-4 & 0.003 & 19.39 & 2.0766 & 2.1241\\
(3,10) & 2.1123(52) & 0.8748(85) & 2-6 & 0.116 & 20.39 & 2.1250 & 2.2314\\
(3,11) & 2.1658(57) & 0.8698(92) & 2-7 & 0.241 & 21.39 & 2.1731 & 2.3377\\
(3,12) & 2.2234(65) & 0.8720(105) & 2-5 & 0.373 & 22.39 & 2.2208 & 2.4431\\
(3,13) & 2.2698(73) & 0.8548(118) & 2-5 & 0.168 & 23.39 & 2.2683 & 2.5478\\
(3,14) & 2.3192(88) & 0.8418(143) & 2-5 & 0.417 & 24.39 & 2.3157 & 2.6521\\
(3,15) & 2.3843(94) & 0.8564(154) & 2-6 & 0.653 & 25.39 & 2.3628 & 2.7559\\
(3,16) & 2.4393(112) & 0.8530(185) & 2-5 & 0.469&26.39 & 2.4098 & 2.8594\\
(4,1) & 0.8228(21) & 0.3069(34) & 5-12 & 0.586 & 16.12 & 1.8119 & 0.8224\\
(4,2) & 1.2510(77) & 0.3541(110) & 4-9 & 0.528 & 16.49 & 1.8975 & 1.2004\\
(4,3) & 1.4643(247) & 0.3390(336) & 4-7 & 0.247 & 17.09&1.9482 & 1.3939\\
(4,4) & 1.7452(112) & 0.5783(194) & 3-7 & 0.910 & 17.89&1.9972 & 1.5412\\
(4,5) & 1.9147(159) & 0.7137(337) & 3-5 & 1.694 & 18.86&2.0493 & 1.6701\\
(4,6) & 2.0168(230) & 0.7591(518) & 3-7 & 0.425 & 19.86&2.1007 & 1.7897\\
(4,7) & 2.1167(49) & 0.8435(84) & 2-4 & 0.041 & 20.86 & 2.1507 & 1.9041\\
(4,8) & 2.1712(56) & 0.8341(89) & 2-4 & 0.436 & 21.86 & 2.2000 & 2.0152\\
(4,9) & 2.2444(62) & 0.8611(102) & 2-6 & 0.363 & 22.86 & 2.2486 & 2.1241\\
(4,10) & 2.2970(84) & 0.8505(138) & 2-5 & 1.626 & 23.86 & 2.2968 & 2.2314\\
(4,11) & 2.3430(84) & 0.8332(133) & 2-6 & 0.514 & 24.86 & 2.3447 & 2.3377\\
(4,12) & 2.3976(96) & 0.8309(149) & 2-5 & 0.544 & 25.86 & 2.3923 & 2.4431\\
(4,13) & 2.4503(109) & 0.8225(168) & 2-4 & 0.030&26.86 & 2.4397 & 2.5478\\
(4,14) & 2.5043(123) & 0.8178(194) & 2-4 & 0.070&27.86 & 2.4869 & 2.6521\\
(4,16) & 2.6037(177) & 0.7952(276) & 2-4 & 2.015&29.86 & 2.5809 & 2.8594\\
\end{tabular}
\end{ruledtabular}
\end{table*}

\begin{table*}[t]
\caption{
A part of lattice QCD results of the 4Q potential $V_{\rm 4Q}$ 
for the twisted 4Q configuration as shown in Fig.7 with $d_1=d_2=d_3=d_4\equiv d$. 
The symmetric twisted 4Q systems are labeled by $(d,h)$.
The notations are the same as in Table~I.}
\label{Table3}
\begin{ruledtabular}
\begin{tabular}{cccccccc} 
($d,h$) & $V_{\rm 4Q}$     & $\bar C$ & $T_{\rm min}-T_{\rm max}$ 
& $\chi^2/N_{\rm DF}$ 
& $L_{\rm min}^{\rm c4Q}$
& $V_{\rm c4Q}^{\rm th}$ 
& $V_{\rm 2Q\bar{Q}}^{\rm th}$\\ \hline
(1,1) & 1.1779(06) & 0.9695(11) & 2-8 & 0.695 &  4.47 & 1.1693 & 1.1305\\
(1,2) & 1.2577(06) & 0.9687(11) & 2-5 & 0.009 &  5.46 & 1.2611 & 1.2967\\
(1,3) & 1.3311(08) & 0.9676(15) & 2-8 & 0.465 &  6.46 & 1.3362 & 1.4435\\
(1,4) & 1.3960(11) & 0.9642(20) & 2-6 & 0.676 &  7.46 & 1.4002 & 1.5737\\
(1,5) & 1.4532(13) & 0.9546(23) & 2-8 & 0.546 &  8.46 & 1.4580 & 1.6941\\
(1,6) & 1.5100(35) & 0.9497(99) & 3-8 & 0.221 &  9.46 & 1.5123 & 1.8088\\
(1,7) & 1.5661(18) & 0.9472(32) & 2-7 & 0.305 & 10.46 & 1.5644 & 1.9200\\
(1,8) & 1.6177(21) & 0.9357(37) & 2-7 & 0.311 & 11.46 & 1.6150 & 2.0288\\
(1,9) & 1.6712(24) & 0.9300(41) & 2-7 & 0.416 & 12.46 & 1.6646 & 2.1360\\
(1,10) & 1.7144(68) & 0.8956(180) & 3-5 & 0.006 & 13.46 & 1.7136 & 2.2421\\
(1,11) & 1.7751(32) & 0.9134(53) & 2-5 & 0.890 & 14.46 & 1.7620 & 2.3472\\
(1,12) & 1.8302(38) & 0.9109(65) & 2-6 & 0.039 & 15.46 & 1.8100 & 2.4518\\
(1,13) & 1.8778(45) & 0.8946(76) & 2-6 & 0.689 & 16.46 & 1.8577 & 2.5558\\
(1,14) & 1.9306(49) & 0.8879(79) & 2-6 & 0.543 & 17.46 & 1.9052 & 2.6595\\
(1,15) & 1.9860(56) & 0.8856(90) & 2-6 & 0.649 & 18.46 & 1.9525 & 2.7628\\
(1,16) & 2.0378(62) & 0.8771(99) & 2-5 & 0.100 & 19.46 & 1.9996 & 2.8658\\
(2,1) & 1.4571(27) & 0.9244(72) & 3-6 & 0.130 &  8.25 & 1.4778 & 1.3939\\
(2,2) & 1.5027(32) & 0.9296(89) & 3-8 & 0.131 &  8.94 & 1.5221 & 1.4656\\
(2,3) & 1.5613(14) & 0.9445(24) & 2-4 & 0.033 &  9.93 & 1.5800 & 1.5578\\
(2,4) & 1.6152(18) & 0.9388(31) & 2-7 & 0.339 & 10.93 & 1.6365 & 1.6576\\
(2,5) & 1.6713(18) & 0.9367(31) & 2-4 & 0.499 & 11.93 & 1.6907 & 1.7598\\
(2,6) & 1.7249(22) & 0.9305(38) & 2-4 & 0.266 & 12.93 & 1.7431 & 1.8626\\
(2,7) & 1.7780(27) & 0.9241(47) & 2-7 & 0.439 & 13.93 & 1.7941 & 1.9655\\
(2,8) & 1.8305(29) & 0.9162(50) & 2-4 & 0.044 & 14.93 & 1.8441 & 2.0683\\
(2,9) & 1.8752(35) & 0.8945(58) & 2-6 & 2.241 & 15.93 & 1.8933 & 2.1708\\
(2,10) & 1.9335(40) & 0.8985(68) & 2-6 & 0.989 & 16.93 & 1.9420 & 2.2732\\
(2,11) & 1.9882(45) & 0.8961(74) & 2-6 & 0.776 & 17.93 & 1.9902 & 2.3755\\
(2,12) & 2.0295(51) & 0.8681(82) & 2-5 & 1.912 & 18.93 & 2.0381 & 2.4776\\
(2,13) & 2.0933(56) & 0.8818(92) & 2-4 & 0.042 & 19.93 & 2.0857 & 2.5796\\
(2,14) & 2.1470(62) & 0.8768(102) & 2-5 & 0.015 & 20.93 & 2.1331 & 2.6815\\
(2,15) & 2.1910(70) & 0.8552(115) & 2-7 & 0.874 & 21.93 & 2.1804 & 2.7833\\
(2,16) & 2.2315(76) & 0.8272(119) & 2-5 & 0.884 & 22.93 & 2.2274 & 2.8850\\
\end{tabular}
\end{ruledtabular}
\end{table*}

\begin{table*}[t]
\caption{
A part of lattice QCD results of the 4Q potential $V_{\rm 4Q}$ 
for the symmetric twisted 4Q configuration labeled by $(d,h)$.
The notations are the same as in Table~I.}
\label{Table4}
\begin{ruledtabular}
\begin{tabular}{cccccccc} 
($d,h$) & $V_{\rm 4Q}$     & $\bar C$ & $T_{\rm min}-T_{\rm max}$ 
& $\chi^2/N_{\rm DF}$
& $L_{\rm min}^{\rm c4Q}$ 
& $V_{\rm c4Q}^{\rm th}$ 
& $V_{\rm 2Q\bar{Q}}^{\rm th}$\\ \hline
(3,1) & 1.6641(20) & 0.9086(34) & 2-8 & 1.052 & 12.17 & 1.7093 & 1.5889\\
(3,2) & 1.6980(21) & 0.9095(35) & 2-4 & 0.227 & 12.65 & 1.7359 & 1.6314\\
(3,3) & 1.7444(25) & 0.9098(43) & 2-6 & 0.649 & 13.42 & 1.7769 & 1.6941\\
(3,4) & 1.7960(26) & 0.9085(46) & 2-5 & 0.278 & 14.39 & 1.8275 & 1.7700\\
(3,5) & 1.8473(31) & 0.9017(53) & 2-5 & 0.747 & 15.39 & 1.8787 & 1.8540\\
(3,6) & 1.9015(36) & 0.8995(60) & 2-6 & 0.151 & 16.39 & 1.9292 & 1.9431\\
(3,7) & 1.9563(39) & 0.8969(66) & 2-6 & 0.334 & 17.39 & 1.9790 & 2.0355\\
(3,8) & 2.0077(46) & 0.8888(75) & 2-6 & 0.216 & 18.39 & 2.0282 & 2.1301\\
(3,9) & 2.0609(47) & 0.8838(77) & 2-6 & 0.067 & 19.39 & 2.0769 & 2.2261\\
(3,10) & 2.1146(52) & 0.8787(86) & 2-5 & 0.121 & 20.39 & 2.1252 & 2.3232\\
(3,11) & 2.1695(56) & 0.8763(91) & 2-4 & 0.104 & 21.39 & 2.1732 & 2.4210\\
(3,12) & 2.2284(65) & 0.8805(109) & 2-6 & 0.053 & 22.39 & 2.2209 & 2.5194\\
(3,13) & 2.2684(69) & 0.8513(112) & 2-6 & 0.676 & 23.39 & 2.2684 & 2.6182\\
(3,14) & 2.3303(82) & 0.8597(133) & 2-4 & 0.373 & 24.39 & 2.3157 & 2.7174\\
(3,15) & 2.3725(96) & 0.8357(151) & 2-5 & 0.631 & 25.39 & 2.3629 & 2.8168\\
(3,16) & 2.4051(112) & 0.7968(168) & 2-6 & 0.378&26.39 & 2.4099 & 2.9165\\
(4,1) & 1.8453(33) & 0.8666(56) & 2-6 & 0.389 & 16.12 & 1.9179 & 1.7598\\
(4,2) & 1.8396(124) & 0.7745(285) & 3-6 & 1.225 & 16.49&1.9368 & 1.7897\\
(4,3) & 1.8832(135) & 0.7758(315) & 3-6 & 1.204 & 17.09&1.9671 & 1.8363\\
(4,4) & 1.9745(44) & 0.8706(71) & 2-4 & 0.193 & 17.89 & 2.0072 & 1.8960\\
(4,5) & 2.0271(47) & 0.8670(79) & 2-6 & 0.321 & 18.86 & 2.0549 & 1.9655\\
(4,6) & 2.0749(52) & 0.8536(85) & 2-5 & 0.226 & 19.86 & 2.1040 & 2.0422\\
(4,7) & 2.1306(52) & 0.8522(86) & 2-4 & 0.543 & 20.86 & 2.1528 & 2.1241\\
(4,8) & 2.1932(66) & 0.8640(108) & 2-6 & 0.250 & 21.86 & 2.2012 & 2.2099\\
(4,9) & 2.2409(62) & 0.8510(101) & 2-5 & 0.801 & 22.86 & 2.2494 & 2.2985\\
(4,10) & 2.2867(66) & 0.8331(107) & 2-4 & 0.060 & 23.86 & 2.2973 & 2.3893\\
(4,11) & 2.3324(87) & 0.8152(135) & 2-4 & 0.680 & 24.86 & 2.3450 & 2.4818\\
(4,12) & 2.3951(107) & 0.8247(166) & 2-5 & 0.199&25.86 & 2.3925 & 2.5756\\
(4,13) & 2.4477(110) & 0.8173(174) & 2-6 & 2.138&26.86 & 2.4398 & 2.6705\\
(4,14) & 2.4910(128) & 0.7958(196) & 2-5 & 0.053&27.86 & 2.4870 & 2.7662\\
(4,15) & 2.5625(154) & 0.8210(238) & 2-6 & 0.904&28.86 & 2.5341 & 2.8626\\
(4,16) & 2.5995(155) & 0.7897(238) & 2-5 & 0.099&29.86 & 2.5810 & 2.9596\\
\end{tabular}
\end{ruledtabular}
\end{table*}

\begin{table*}[t]
\caption{
Lattice QCD results for the 4Q potential $V_{\rm 4Q}$ for 
the asymmetric planar 4Q configuration 
as shown in Fig.6 with various $(d_1,d_2,d_3,d_4)$ for $h$=8. 
The notations are the same as in Table~I.}
\label{Table5}
\begin{ruledtabular}
\begin{tabular}{cccccccc} 
$(d_1,d_2,d_3,d_4)$ & $V_{\rm 4Q}$     & $\bar C$ & $T_{\rm min}-T_{\rm max}$ 
& $\chi^2/N_{\rm DF}$
& $L_{\rm min}^{\rm c4Q}$ 
& $V_{\rm c4Q}^{\rm th}$ 
& $V_{\rm 2Q\bar{Q}}^{\rm th}$\\ \hline
(0,1,0,1) & 1.4227(18) & 0.9368(31) & 2-10 & 0.610 &  9.73 & 1.3983 & 2.0152\\
(0,1,1,0) & 1.4243(17) & 0.9377(29) & 2-10 & 0.265 &  9.78 & 1.4008 & 2.0220\\
(0,1,1,1) & 1.5215(19) & 0.9385(34) & 2-8 & 0.544 & 10.61 & 1.5073 & 2.0186\\
(0,1,2,0) & 1.5170(39) & 0.9054(103) & 3-7 & 0.424 & 10.70 & 1.5118 & 2.0321\\
(0,1,1,2) & 1.5689(43) & 0.8991(115) & 3-8 & 0.596 & 11.46 & 1.5697 & 2.0220\\
(0,1,2,1) & 1.5830(21) & 0.9353(36) & 2-7 & 0.155 & 11.51 & 1.5718 & 2.0287\\
(0,1,3,0) & 1.5926(21) & 0.9289(36) & 2-7 & 0.086 & 11.64 & 1.5781 & 2.0483\\
(0,2,1,1) & 1.6211(21) & 0.9300(38) & 2-7 & 0.388 & 11.51 & 1.6171 & 2.0220\\
(0,2,2,0) & 1.6301(22) & 0.9310(40) & 2-7 & 0.199 & 11.64 & 1.6234 & 2.0422\\
(0,1,1,3) & 1.6281(22) & 0.9200(38) & 2-8 & 0.687 & 12.34 & 1.6219 & 2.0321\\
(0,1,2,2) & 1.6231(49) & 0.9144(132) & 3-8 & 0.221 & 12.34 & 1.6219 & 2.0321\\
(0,1,3,1) & 1.6326(23) & 0.9253(40) & 2-7 & 0.758 & 12.42 & 1.6258 & 2.0449\\
(0,2,1,2) & 1.6809(22) & 0.9262(39) & 2-6 & 0.779 & 12.34 & 1.6784 & 2.0186\\
(0,2,2,1) & 1.6834(23) & 0.9273(42) & 2-8 & 0.762 & 12.42 & 1.6824 & 2.0321\\
(0,2,3,0) & 1.6934(23) & 0.9208(39) & 2-5 & 0.470 & 12.58 & 1.6902 & 2.0584\\
(1,1,1,2) & 1.6784(23) & 0.9324(41) & 2-7 & 0.510 & 12.34 & 1.6784 & 2.0186\\
(0,1,2,3) & 1.6668(61) & 0.9034(165) & 3-6 & 0.309 & 13.20 & 1.6687 & 2.0422\\
(0,1,3,2) & 1.6732(25) & 0.9199(45) & 2-6 & 0.502 & 13.23 & 1.6705 & 2.0483\\
(0,2,1,3) & 1.7270(25) & 0.9125(45) & 2-6 & 0.683 & 13.20 & 1.7296 & 2.0220\\
(0,2,3,1) & 1.7349(25) & 0.9210(43) & 2-7 & 0.323 & 13.35 & 1.7370 & 2.0483\\
(0,3,1,2) & 1.7441(26) & 0.9177(47) & 2-7 & 0.369 & 13.23 & 1.7426 & 2.0220\\
(0,3,2,1) & 1.7473(26) & 0.9191(46) & 2-7 & 0.503 & 13.35 & 1.7482 & 2.0422\\
(0,3,3,0) & 1.7604(25) & 0.9172(45) & 2-7 & 0.166 & 13.53 & 1.7573 & 2.0747\\
(1,1,1,3) & 1.7295(26) & 0.9259(45) & 2-7 & 0.880 & 13.23 & 1.7314 & 2.0287\\
(1,1,2,2) & 1.7218(24) & 0.9221(45) & 2-6 & 0.307 & 13.20 & 1.7296 & 2.0220\\
(1,2,1,2) & 1.7398(24) & 0.9311(44) & 2-6 & 0.254 & 13.20 & 1.7408 & 2.0152\\
(1,2,2,1) & 1.7299(66) & 0.9013(176) & 3-9 & 0.298 & 13.23 & 1.7426 & 2.0220\\
(0,1,3,3) & 1.7159(29) & 0.9127(49) & 2-6 & 0.559 & 14.07 & 1.7142 & 2.0584\\
(0,2,3,2) & 1.7754(29) & 0.9167(49) & 2-9 & 0.893 & 14.14 & 1.7807 & 2.0449\\
(0,3,3,1) & 1.8006(28) & 0.9162(49) & 2-6 & 1.013 & 14.28 & 1.8033 & 2.0584\\
(1,1,2,3) & 1.7687(27) & 0.9179(47) & 2-9 & 0.558 & 14.07 & 1.7772 & 2.0321\\
(1,2,2,2) & 1.7866(27) & 0.9268(49) & 2-6 & 0.192 & 14.07 & 1.7928 & 2.0186\\
(1,2,3,1) & 1.7900(28) & 0.9208(49) & 2-6 & 0.436 & 14.14 & 1.7963 & 2.0321\\
(1,2,2,3) & 1.8273(31) & 0.9109(55) & 2-6 & 0.881 & 14.93 & 1.8395 & 2.0220\\
(1,2,3,2) & 1.8311(30) & 0.9170(54) & 2-7 & 0.284 & 14.96 & 1.8412 & 2.0287\\
(1,3,3,1) & 1.8437(32) & 0.9191(57) & 2-6 & 0.449 & 15.06 & 1.8506 & 2.0422\\
(1,3,3,2) & 1.8845(35) & 0.9154(59) & 2-6 & 0.173 & 15.87 & 1.8946 & 2.0321\\
(2,2,2,3) & 1.8774(34) & 0.9131(60) & 2-7 & 0.197 & 15.80 & 1.8914 & 2.0186\\
(2,3,2,3) & 1.9213(37) & 0.9050(63) & 2-6 & 0.030 & 16.66 & 1.9380 & 2.0152\\
(2,3,3,3) & 1.9648(39) & 0.8987(65) & 2-6 & 0.603 & 17.53 & 1.9832 & 2.0186\\
\end{tabular}
\end{ruledtabular}
\end{table*}

\begin{table*}[t]
\caption{
Lattice QCD results for the 4Q potential $V_{\rm 4Q}$ for 
the asymmetric twisted 4Q configuration 
as shown in Fig.7 with various $(d_1,d_2,d_3,d_4)$ for $h$=8. 
The notations are the same as in Table~I.}
\label{Table6}
\begin{ruledtabular}
\begin{tabular}{cccccccc} 
$(d_1,d_2,d_3,d_4)$ & $V_{\rm 4Q}$     & $\bar C$ & $T_{\rm min}-T_{\rm max}$ 
& $\chi^2/N_{\rm DF}$
& $L_{\rm min}^{\rm c4Q}$ 
& $V_{\rm c4Q}^{\rm th}$ 
& $V_{\rm 2Q\bar{Q}}^{\rm th}$\\ \hline
(0,1,0,1) & 1.4179(34) & 0.9197(92) & 3-8 & 0.857 &  9.76 & 1.3997 & 2.0220\\
(0,1,1,1) & 1.5239(18) & 0.9435(34) & 2-10 & 0.590 & 10.61 & 1.5074 & 2.0254\\
(0,1,0,2) & 1.5152(129) & 0.8961(453) & 4-9 & 0.173 & 10.66 & 1.5100 & 2.0320\\
(0,1,1,2) & 1.5819(47) & 0.9340(132) & 3-7 & 0.262 & 11.49 & 1.5709 & 2.0354\\
(0,1,0,3) & 1.5896(21) & 0.9257(37) & 2-9 & 0.525 & 11.58 & 1.5756 & 2.0481\\
(0,2,1,1) & 1.6248(20) & 0.9373(36) & 2-5 & 0.889 & 11.52 & 1.6176 & 2.0354\\
(0,2,0,2) & 1.6259(22) & 0.9263(38) & 2-7 & 0.367 & 11.57 & 1.6202 & 2.0417\\
(0,1,1,3) & 1.6279(53) & 0.9144(143) & 3-7 & 0.538 & 12.39 & 1.6241 & 2.0515\\
(0,1,2,2) & 1.6252(50) & 0.9204(140) & 3-9 & 0.070 & 12.34 & 1.6219 & 2.0455\\
(0,2,1,2) & 1.6813(24) & 0.9250(42) & 2-8 & 0.770 & 12.39 & 1.6811 & 2.0452\\
(0,2,0,3) & 1.6923(23) & 0.9233(38) & 2-7 & 0.545 & 12.49 & 1.6858 & 2.0575\\
(1,1,1,2) & 1.6790(22) & 0.9334(39) & 2-7 & 0.652 & 12.34 & 1.6785 & 2.0388\\
(0,1,2,3) & 1.6724(24) & 0.9192(43) & 2-8 & 0.221 & 13.22 & 1.6696 & 2.0616\\
(0,2,1,3) & 1.7335(25) & 0.9213(45) & 2-7 & 0.040 & 13.29 & 1.7343 & 2.0609\\
(0,3,1,2) & 1.7462(25) & 0.9187(45) & 2-7 & 0.323 & 13.31 & 1.7467 & 2.0609\\
(0,3,0,3) & 1.7567(27) & 0.9166(47) & 2-6 & 0.557 & 13.41 & 1.7514 & 2.0726\\
(1,1,1,3) & 1.7293(25) & 0.9252(45) & 2-7 & 0.392 & 13.24 & 1.7317 & 2.0549\\
(1,1,2,2) & 1.7220(24) & 0.9222(45) & 2-6 & 0.420 & 13.20 & 1.7296 & 2.0488\\
(1,2,1,2) & 1.7388(27) & 0.9280(46) & 2-6 & 0.721 & 13.22 & 1.7420 & 2.0485\\
(0,1,3,3) & 1.7198(27) & 0.9199(47) & 2-8 & 0.746 & 14.08 & 1.7141 & 2.0778\\
(0,2,2,3) & 1.7734(28) & 0.9138(49) & 2-7 & 0.497 & 14.12 & 1.7798 & 2.0710\\
(0,3,1,3) & 1.7979(28) & 0.9145(49) & 2-6 & 0.096 & 14.21 & 1.7999 & 2.0761\\
(1,1,2,3) & 1.7687(29) & 0.9174(53) & 2-8 & 0.832 & 14.07 & 1.7772 & 2.0649\\
(1,2,2,2) & 1.7840(28) & 0.9210(49) & 2-7 & 0.820 & 14.07 & 1.7930 & 2.0585\\
(1,2,1,3) & 1.7890(30) & 0.9201(51) & 2-7 & 1.063 & 14.12 & 1.7952 & 2.0643\\
(1,2,2,3) & 1.8312(30) & 0.9169(53) & 2-6 & 0.095 & 14.95 & 1.8407 & 2.0743\\
(1,3,1,3) & 1.8393(31) & 0.9128(54) & 2-7 & 0.474 & 15.02 & 1.8484 & 2.0794\\
(1,3,2,3) & 1.8853(34) & 0.9164(58) & 2-6 & 1.208 & 15.85 & 1.8939 & 2.0894\\
(2,2,2,3) & 1.8807(33) & 0.9180(55) & 2-4 & 0.032 & 15.80 & 1.8917 & 2.0840\\
(2,3,2,3) & 1.9171(36) & 0.8943(64) & 2-5 & 0.564 & 16.68 & 1.9394 & 2.0992\\
(2,3,3,3) & 1.9653(40) & 0.8968(69) & 2-6 & 0.128 & 17.54 & 1.9838 & 2.1149\\
\end{tabular}
\end{ruledtabular}
\end{table*}

\section{Discussions}
\label{discussion}

\subsection{Comparison with theoretical Ans\"atze}

In this section, we compare the lattice QCD results of the 4Q potential $V_{\rm 4Q}$ 
with the theoretical Ans\"atze presented in Sect.II, i.e., 
the OGE plus multi-Y Ansatz (\ref{Vc4Q}) and the two-meson Ansatz (\ref{V2QQbar}). 

For the OGE plus multi-Y Ansatz (\ref{Vc4Q}), 
we set the parameters ($A_{\rm 4Q}, \sigma_{\rm 4Q}$) to be 
($A_{\rm 3Q}, \sigma_{\rm 3Q}$) in the 3Q potential $V_{\rm 3Q}$ in Ref.\cite{TSNM02}, 
i.e., $A_{\rm 4Q}=A_{\rm 3Q} \simeq$ 0.1366,
$\sigma_{\rm 4Q}=\sigma_{\rm 3Q} \simeq$ 0.0460$a^{-2}$.
Note that there are no adjustable parameters 
except for an irrelevant constant $C_{\rm 4Q} \simeq$ 1.2579$a^{-1}$.

For the two-meson Ansatz (\ref{V2QQbar}), 
we adopt the lattice result for the Q$\rm \bar{Q}$ potential $V_{\rm Q \bar Q}$ 
in Ref.\cite{TSNM02}, i.e., 
$A_{\rm Q\bar{Q}} \simeq$ 0.2768, 
$\sigma_{\rm Q\bar{Q}} \simeq$ 0.0506$a^{-2}$, 
$C_{\rm Q\bar{Q}} \simeq$ 0.6374$a^{-1}$. 
Then, there are no adjustable parameters also for the two-meson Ansatz.

We demonstrate the two Ans\"atze 
for the symmetric planar 4Q configurations 
as shown in Fig.6 with $d_1=d_2=d_3=d_4 \equiv d$. 
In this case, the OGE Coulomb plus multi-Y Ansatz for the connected 4Q system reads 
\begin{eqnarray}
V_{\rm c4Q}(d,h)&=&-A_{\rm 4Q}\left(\frac{1}{d}+\frac{1}{h}+\frac{1}{\sqrt{h^2+4d^2}}\right) \nonumber \\
   &+&\sigma_{\rm 4Q}L_{\rm min} +C_{\rm 4Q}. 
\label{Vc4Qpl}
\end{eqnarray}
In the case of $h \ge \frac{2}{\sqrt{3}}d$, 
the lowest connected 4Q system takes the double-Y-shaped flux-tube, and  
the minimal value of the total flux-tube length is expressed as 
\begin{eqnarray}
L_{\rm min}=h+2\sqrt{3} d.
\end{eqnarray} 
In the case of $h \le \frac{2}{\sqrt{3}}d$, 
the lowest connected 4Q system takes the X-shaped flux-tube with $L_{\rm min}=2\sqrt{h^2+4d^2}$, 
although it must be unstable against the decay into two mesons. 
On the other hand, the two-meson Ansatz reads  
\begin{eqnarray}
V_{\rm 2Q\bar Q}(h)=2 \times V_{\rm Q\bar Q}(h), 
\label{V2QQbarpl}
\end{eqnarray}
which is independent of $d$.

We show in Fig.10(a) the lattice QCD results of the 4Q potential $V_{\rm 4Q}$ 
for symmetric planar 4Q configurations \cite{STOI04b,OST04p,AK05} 
with $d=1-4$.
The symbols denote lattice QCD results.
The curves describe the theoretical form:
the solid curve denotes the OGE plus multi-Y Ansatz~(\ref{Vc4Q}), and 
the dashed-dotted curve the two-meson Ansatz~(\ref{V2QQbar}).

For large value of $h$ compared with $d$, the lattice data seem to coincide with 
the OGE Coulomb plus multi-Y Ansatz \cite{STOI04b,OST04p}.
On the other hand, for small $h$, the lattice data tend to agree with the two-meson Ansatz and 
seem independent of $d$ \cite{STOI04b,OST04p}.
These tendencies were also observed in a recent lattice work by other group \cite{AK05}.
This would correspond to the transition from the connected 4Q state into the two-meson state as $h$ decreases, 
as will be discussed in the next subsection.

Here, we comment on the transition in terms of the ground-state overlap $\bar C$ listed in Table~I.
For large $h$, the ground-state overlap $\bar C$ is almost unity, 
which implies the realization of the quasi-ground-state in  the present calculation with 
the smeared 4Q Wilson loop based on the connected 4Q configuration.
For small $h$, however, $\bar C$ tends to be small, 
and hence, for accurate measurements, we have to take relatively large values of $T$ as the fit range. 
This would indicate that the ground-state configuration is largely different from the connected 4Q configuration for small $h$.
(In other words, it may be nontrivial to obtain the result indicating the two-meson state for small $h$ 
from the 4Q Wilson loop based on the connected 4Q configuration.)

\begin{figure}[ht]
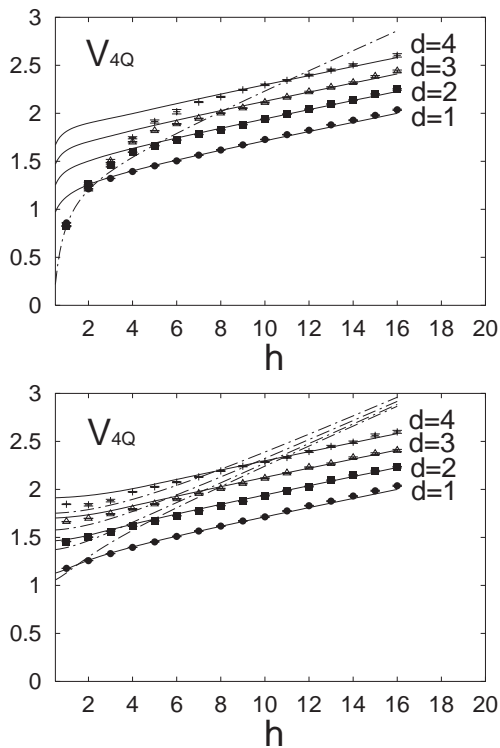

\begin{center}
\includegraphics[width=6.5cm]{Fig10a.eps}
\includegraphics[width=6.5cm]{Fig10b.eps}
\end{center}
\caption
{The tetraquark (4Q) potential $V_{\rm 4Q}$: 
(a) for symmetric planar 4Q configurations as shown in Fig.6; 
(b) for symmetric twisted 4Q configurations as shown in Fig.7.
The symbols denote the lattice QCD results.
The curves describe the theoretical form:
the solid curve denotes the OGE plus multi-Y Ansatz, and 
the dashed-dotted curve the two-meson Ansatz.
}
\label{Fig10}
\end{figure}

Next, we investigate the twisted 4Q configuration 
\cite{STOI04b,OST04p} as shown in Fig.7.
We show in Fig.10(b) the lattice QCD results of the 4Q potential
for symmetric twisted 4Q configurations with $d=1-4$.
The symbols denote lattice QCD results for each $d$, 
and the curves describe 
the theoretical form of the OGE plus multi-Y Ansatz.

The lattice data seem to agree with the OGE plus multi-Y Ansatz 
in the wide region of $h$ \cite{STOI04b,OST04p}.
In the twisted 4Q configuration, the distance between the nearest quark and 
antiquark cannot take a smaller value than the ``inter-diquark distance" $h$, 
and therefore $V_{\rm c4Q}$ is smaller than $V_{\rm 2Q\bar Q}$ in most cases 
except for extreme configurations as $d>h$. (See Fig.10(b).)
Then, different from the planar case, it is not easy to make the transition from 
the connected 4Q state into the two-meson state for the twisted case. 
Also from the lattice data, the ground-state overlap $\bar C$ is found to be almost unity for all twisted 4Q configurations, 
which indicates that the ground state resembles a connected 4Q state.
In other words, the twisted 4Q configuration seems to be rather stable against the transition into the two $\rm Q\bar Q$ mesons, 
which may indicate a stability of the ``twisted structure" or the ``tetrahedral structure" of the 4Q system. 

We also investigate more general asymmetric 4Q configurations 
with various $(d_1,d_2,d_3,d_4)$ 
for both planar and twisted cases, as shown in Table V and VI.
Also for the asymmetric planar and twisted 4Q configurations, 
$V_{\rm 4Q}$ seems to be well described with the OGE plus multi-Y Ansatz 
in the case of $h > \frac{2}{\sqrt{3}}d_i$ $(i=1,2,3,4)$.
Note here that some 4Q configurations are physically equivalent, 
e.g., the planar cases with $(d_1,d_2,d_3,d_4)$=(1,1,1,2) and (0,2,1,2), 
although the corresponding smeared 4Q Wilson loops are different. 
For such cases, the lattice QCD results are found to be almost the same.  
In fact, the extracted lattice results are 
almost independent of the way how the 4Q Wilson loop is constructed, 
as long as the spatial locations of the static four quarks are the same.
This indicates that the ground-state contribution is properly extracted 
in the present calculation.

As the conclusion, 
the OGE plus multi-Y Ansatz well describes the 4Q potential $V_{\rm 4Q}$, 
when QQ and $\rm \bar{Q}\bar{Q}$ are well separated, e.g., 
the ``inter-diquark distance" $h$ is large in comparison with the ``diquark size" $d$.
On the other hand, when the nearest quark and antiquark pair is spatially close, 
the system is described as a ``two-meson" state.

Together with the previous studies \cite{TMNS01,TSNM02,OST04,STOI04a,STOI04b,SOTI04,SIOT04,OST04p} 
for the inter-quark potentials in lattice QCD,
we have found the universality of the string tension as 
\begin{eqnarray}
\sigma_{\rm Q\bar{Q}} \simeq \sigma_{\rm 3Q} \simeq 
\sigma_{\rm 4Q} \simeq \sigma_{\rm 5Q} \simeq (420{\rm MeV})^2,
\end{eqnarray}
and the OGE result for the Coulomb coefficient as 
\begin{eqnarray}
A_{\rm Q\bar{Q}} \simeq 2A_{\rm 3Q} \simeq 
2A_{\rm 4Q} \simeq 2A_{\rm 5Q} \simeq 0.27.
\end{eqnarray}
In particular, these lattice QCD studies \cite{OST04,STOI04a,STOI04b,SOTI04,SIOT04,OST04p} 
indicate a fairly good agreement among 
$\sigma_{\rm 3Q}$, $\sigma_{\rm 4Q}$ and $\sigma_{\rm 5Q}$, 
which seem to be slightly smaller than $\sigma_{\rm Q\bar Q}$. 
(As an interesting possibility, the numerical similarity 
among $\sigma_{\rm 3Q}$, $\sigma_{\rm 4Q}$ and $\sigma_{\rm 5Q}$ 
may reflect the similar structure of the Y-type flux-tube in the multi-quark systems.)
The universality of the string tension observed in our lattice QCD studies 
\cite{TMNS01,TSNM02,OST04,STOI04a,STOI04b,SOTI04,SIOT04,OST04p} 
seems to be consistent with the hypothetical flux-tube picture \cite{N74,KS75,CKP83,CNN79,IP8385,P84,O85,LLMRSY86}
or the strong-coupling expansion scheme \cite{KS75,CKP83}, 
although strong-coupling QCD does not have a continuum limit and is far from real QCD.
As for the irrelevant constant, 
$C_{\rm Q\bar Q}$, $C_{\rm 3Q}$, $C_{\rm 4Q}$ and $C_{\rm 5Q}$ 
are non-scaling unphysical quantities appearing in the lattice regularization, 
and we find an approximate relation as 
\begin{eqnarray}
\frac{C_{\rm Q\bar Q}}{2} \simeq \frac{C_{\rm 3Q}}{3} 
\simeq \frac{C_{\rm 4Q} }{4} \simeq \frac{C_{\rm 5Q} }{5} \simeq 0.32 a^{-1}
\end{eqnarray}
in lattice QCD \cite{TMNS01,TSNM02,OST04,STOI04a,STOI04b,SOTI04,SIOT04,OST04p}.

\subsection{The quark confinement force in 4Q systems}

While the short-distance OGE Coulomb force can be understood with perturbative QCD, 
the long-distance confinement force is a typical nonperturbative quantity 
and highly nontrivial particularly for multi-quark systems. 
To specify the long-distance property of $V_{\rm 4Q}$ 
is important to clarify the confinement mechanism 
from a wide viewpoint including multi-quarks, 
and it also leads to 
a proper quark-model Hamiltonian to describe multi-quark systems.
Therefore, we perform a further analysis for the long-distance force in 4Q systems.

To clarify the long-distance force in the 4Q system, 
we plot $V_{\rm 4Q}$ against $L_{\rm min}^{\rm c4Q}$
for planar and twisted 4Q configurations in Figs.11(a) and (b), respectively.
Here, $L_{\rm min}^{\rm c4Q}$ denotes the minimal flux-tube length for 
the connected 4Q system.
In both planar and twisted cases, for large $L_{\rm min}^{\rm c4Q}$, 
$V_{\rm 4Q}$ approaches a linearly arising function of $L_{\rm min}^{\rm c4Q}$.

\begin{figure}[ht]
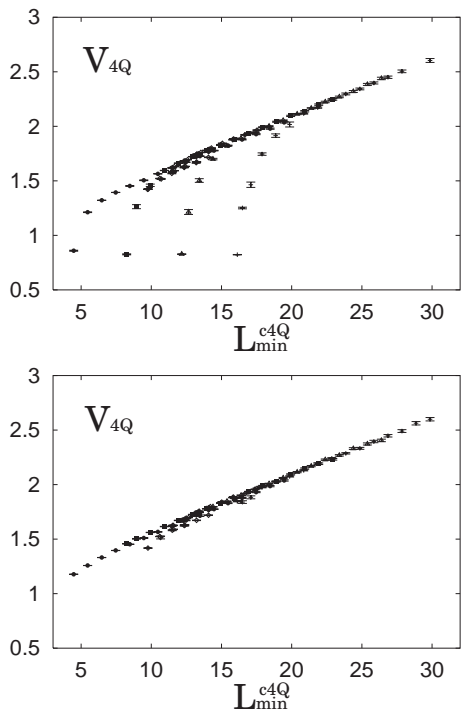

\begin{center}
\includegraphics[width=6cm]{Fig11a.eps}
\includegraphics[width=6cm]{Fig11b.eps}
\end{center}
\caption{
The 4Q potential $V_{\rm 4Q}$ plotted against $L_{\rm min}^{\rm c4Q}$:
(a) for planar 4Q configurations; (b) for twisted 4Q configurations. 
$L_{\rm min}^{\rm c4Q}$ denotes the minimal flux-tube length for the connected 4Q configuration.
We plot all the lattice QCD data of $V_{\rm 4Q}$ 
for the symmetric case with $d=1$ (solid circles),  $d=2$ (solid squares), 
$d=3$ (open triangles) and $d=4$ (crosses) together with 
the asymmetric case (open diamonds). 
}
\label{Fig11}
\end{figure}

To single out the long-distance confinement force, 
we consider the 4Q potential subtracted by the Coulomb part.
Here, we subtract the OGE Coulomb part $V_{\rm c4Q}^{\rm Coul}$ 
of $V_{\rm c4Q}$ in Eq.(\ref{Vc4Q}) for the connected 4Q system, 
with the Coulomb coefficient $A_{\rm 4Q}$ fixed to be 
$A_{\rm 3Q}$ in the 3Q potential $V_{\rm 3Q}$ in Ref.\cite{TSNM02}.
We plot $V_{\rm 4Q}-V_{\rm c4Q}^{\rm Coul}$ against $L_{\rm min}^{\rm c4Q}$ 
for planar and twisted 4Q configurations in Figs.12 (a) and (b), respectively.
For the planar 4Q system, 
$V_{\rm 4Q}-V_{\rm c4Q}^{\rm Coul}$ approaches $\sigma_{\rm 4Q} L_{\rm min}^{\rm c4Q}+C_{\rm 4Q}$ 
except for a small $h$ region, where the flip-flop into a two-meson state can take place. 
For the twisted 4Q system, we observe remarkable agreement between 
the lattice QCD data of $V_{\rm 4Q}-V_{\rm c4Q}^{\rm Coul}$ and 
$\sigma_{\rm 4Q} L_{\rm min}^{\rm c4Q}+C_{\rm 4Q}$ 
for the wide region of $L_{\rm min}$, 
which corresponds that the flip-flop into the two-meson state 
does not occur in most twisted 4Q configurations.

\begin{figure}[ht]
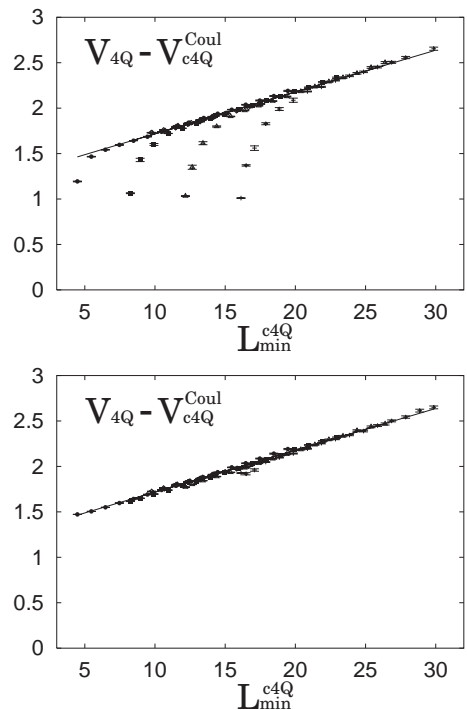

\begin{center}
\includegraphics[width=6cm]{Fig12a.eps}
\includegraphics[width=6cm]{Fig12b.eps}
\end{center}
\caption{
The 4Q potential subtracted by the OGE Coulomb part of the connected 4Q system, 
$V_{\rm 4Q}-V_{\rm c4Q}^{\rm Coul}$, plotted against $L_{\rm min}^{\rm c4Q}$:
(a) for planar 4Q configurations; 
(b) for twisted 4Q configurations.
We plot all the lattice QCD data including asymmetric cases.
The meaning of the symbols is the same as in Fig.11. 
The solid line denotes 
$\sigma_{\rm 4Q} L_{\rm min}^{\rm c4Q}+C_{\rm 4Q}$.
}
\label{Fig12}
\end{figure}

Thus, the confinement potential in the 4Q system as shown in Fig.2
is proportional to $L_{\rm min}$, which indicates that  
the quark confinement force is genuinely 4-body and 
the flux tube is multi-Y shaped. 

\subsection{The flip-flop, the level crossing, and absence of the color van der Waals force}

Finally, we investigate the flip-flop between the connected 4Q state and the two-meson state.
Since the flux-tube changes its shape so as to have the minimal length, 
the multi-Y type flux-tube is expected to change into a two-meson state for small $h$.

This type of the flip-flop is physically important for 
the properties of 4Q states especially for their decay process into two mesons. 
Note also that, in the flux-tube picture, 
the meson-meson reaction is described by 
the flux-tube recombination between the two mesons, 
and this process can be realized through the two successive flip-flops 
between the two-meson state and the connected 4Q state.
Therefore, this type of the flip-flop is important  
also for the reaction mechanism between two mesons. 

As a clear signal of the flip-flop, 
we again show the 4Q potential $V_{\rm 4Q}$ for 
the symmetric planar 4Q configuration 
with $d$=1,2,3 separately in Fig.13.
The solid curve denotes $V_{\rm c4Q}$ for the OGE plus multi-Y Ansatz, and 
the dashed-dotted curve $V_{\rm 2Q\bar Q}=2V_{\rm Q\bar Q}(h)$ for the two-meson Ansatz.
For large $h$, $V_{\rm 4Q}$ coincides with the energy $V_{\rm c4Q}$ of the connected 4Q system. 
For small $h$, $V_{\rm 4Q}$ coincides with the energy $2V_{\rm Q\bar Q}$ of the ``two-meson" system composed of two flux-tubes.
In the intermediate region of $h$, one can observe the cross over from one Ansatz to another.

Thus, in these particular cases, we can observe a clear evidence 
of the flip-flop as 
\begin{eqnarray}
V_{\rm 4Q}(d,h) \simeq {\rm min}(V_{\rm c4Q}(d,h), 2V_{\rm Q\bar Q}(h)),
\end{eqnarray}
which indicates 
the transition between the connected 4Q state and the two-meson state around 
the level-crossing point where these two systems are degenerate 
as $V_{\rm c4Q}(d,h)=2V_{\rm Q\bar Q}(h)$. 
This result also supports the flux-tube picture even for the 4Q system.

The present lattice QCD results on the ``flip-flop" leads to 
infrared screening and disappearance of 
the long-range color interactions, i.e., the confining force and the OGE Coulomb force, 
between (anti)quarks belonging to different ``mesons". 
This physically results in the absence of the tree-level color van der Waals force 
between two mesons \cite{AF78,GYOPRS79,FS79}.

\begin{figure}[ht]
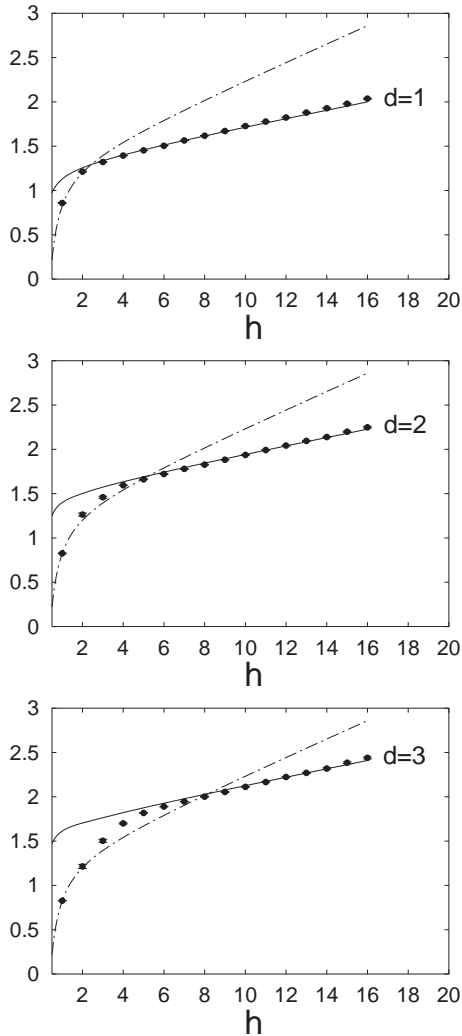

\begin{center}
\includegraphics[width=6cm]{Fig13a.eps}
\includegraphics[width=6cm]{Fig13b.eps}
\includegraphics[width=6cm]{Fig13c.eps}
\end{center}
\caption
{The typical lattice QCD results for the flip-flop between the connected 4Q state
and the two-meson state for the symmetric planar 4Q configuration with fixed $d$.
The symbols denote lattice QCD results.
The curves describe the theoretical form:
the solid curve denotes the OGE plus multi-Y Ansatz, and 
the dashed-dotted curve the two-meson Ansatz.
}
\label{Fig13}
\end{figure}

\section{Summary and concluding remarks}
\label{summary}

We have performed the detailed study of the tetraquark (4Q) potential 
$V_{\rm 4Q}$ for various QQ-$\rm \bar{Q}\bar{Q}$ systems 
in SU(3) lattice QCD with $\beta=6.0$ and $16^3 \times 32$ at the quenched level. 
For about 200 different patterns of 4Q systems, 
we have extracted $V_{\rm 4Q}$ from the 4Q Wilson loop in 300 gauge configurations, 
with the smearing method to enhance the ground-state component.
We have calculated $V_{\rm 4Q}$ for 
planar, twisted, asymmetric, and large-size 4Q configurations, respectively.
The calculation for large-size 4Q configurations has been done 
by identifying $16^2 \times 32$ as the spatial size and 
16 as the temporal one, 
and the long-distance confinement force has been particularly analyzed in terms of the flux-tube picture. 

When QQ and $\rm \bar{Q}\bar{Q}$ are well separated, 
$V_{\rm 4Q}$ is found to be expressed as 
the sum of the one-gluon-exchange Coulomb term and multi-Y type linear term based on the flux-tube picture.
In this case, all the four quarks are linked by the connected double-Y-shaped flux-tube, 
where the total flux-tube length is minimized. 
On the other hand, when the nearest quark and antiquark pair is spatially close, 
the system is described as a ``two-meson" state rather than the connected 4Q state.

We have observed a flux-tube recombination called as ``flip-flop" 
between the connected 4Q state and the ``two-meson" state 
around the level-crossing point.
This ``flip-flop" leads to infrared screening of the long-range color interactions  
between (anti)quarks belonging to different mesons, and results in 
the absence of the tree-level color van der Waals force between two mesons.

As a next step, it is interesting to investigate the transition in terms of the level crossing between the connected 4Q state 
and the two-meson state through the diagonalization of the correlation matrix with various 4Q states \cite{GLPM96}.
Through the investigation of the excited-state levels of the 4Q system, a realistic picture for 
the reaction mechanism between two mesons may be obtained. 

The dynamical quark effect 
for the flux-tube picture and the flip-flop would be also an interesting subject.
In this context, the string-breaking effect may cause more complicated variation of the transition 
between single $\rm Q\bar Q$ meson and the multi-quark system.

In any case, recent lattice QCD studies begin to shed light on the realistic picture in hadron physics and 
to reveal even the properties of the multi-quark system. 

\begin{acknowledgments}
H.S. was supported in part by a Grant for Scientific 
Research (No.16540236) from the Ministry of Education, 
Culture, Science and Technology, Japan.
T.T.T. was supported by the Japan Society for the Promotion 
of Science. 
The lattice QCD Monte Carlo simulations have been performed 
on NEC-SX5 at Osaka University. 
\end{acknowledgments}

\appendix

\section{OGE Coulomb terms in the 4Q potential}

In the Appendix, we briefly show the derivation of the OGE Coulomb terms in the 4Q potential in Eq.(\ref{Vc4Q})
by calculating $\langle {\rm T}| T_i^aT_j^a|{\rm T} \rangle$ for the tetraqurak (4Q) state $\rm |T \rangle$.

In the quark picture, the 4Q state $\rm |T \rangle$ corresponding to Fig.2 is expressed as 
\begin{eqnarray}
|{\rm T} \rangle 
&=& {\rm |(Q_1Q_2)_{\bar{3}}(\bar{Q}_3\bar{Q}_4)_3 \rangle_1} \nonumber \\
&=& \frac{1}{2\sqrt{3}}
\epsilon^{abc}\epsilon^{ab'c'}
|{\rm Q}_1^b {\rm Q}_2^c \bar{\rm Q}_3^{b'} \bar{\rm Q}_4^{c'} \rangle,
\label{A-one}
\end{eqnarray}
where the indices denote the SU(3) color indices of (anti)quarks.
Here, $\rm |T\rangle $ is normalized as 
\begin{eqnarray}
\langle {\rm T} | {\rm T} \rangle = \frac{1}{12}
\epsilon^{abc}\epsilon^{ab'c'}\epsilon^{def}\epsilon^{de'f'}=1.
\end{eqnarray}

The color matrix factor $T_i^a T_j^a$ in the OGE process  
can be expressed with the Casimir operator $C_2(R)$ as 
\begin{eqnarray}
T_i^a T_j^a 
&=& \frac{1}{2}\{ (T_i^a+T_j^b)^2-(T_i^a)^2-(T_j^a)^2 \} \nonumber\\
&=& \frac{1}{2}\{ C_2(R_{i+j})-C_2(R_i)-C_2(R_j) \} \nonumber\\
&=& \frac{1}{2}C_2(R_{i+j})-\frac{4}{3},
\end{eqnarray}
where $R_{i+j}$ denotes the total SU(3) color representation of the $(i+j)$ system.
Here, $C_2(R_i)=C_2(R_j)=\frac43$ has been used for each (anti)quark belonging to ${\bf 3}$ ($\bar{\bf 3}$). 

In this 4Q system, the two quarks, $\rm Q_1$ and $\rm Q_2$, form the $\bar {\bf 3}$ representation, 
i.e., $C_2(R_{1+2})=C_2(\bar {\bf 3})=\frac{4}{3}$, and then one gets $\langle {\rm T}|T_1^a T_2^a|{\rm T}\rangle=-\frac{2}{3}$.
This type of the Coulomb coefficient between two quarks is the same as that in the 3Q system. 
Similarly, one gets $\langle {\rm T}|T_3^a T_4^a|{\rm T}\rangle=-\frac{2}{3}$ for the two antiquarks, 
$\bar {\rm Q}_3$ and $\bar {\rm Q}_4$.

Next, we consider the Coulomb interaction between the quark and the antiquark.
Owing to the symmetry, we only have to investigate the interaction between $\rm Q_1$ and $\rm \bar Q_3$. 
To this end, 
we rewrite the 4Q state $|\rm T\rangle$ in terms of the irreducible representation fo the $\rm Q_1$+$\rm \bar Q_3$ system.
Since $\rm Q_1$ and $\rm \bar Q_3$ can form the singlet ({\bf 1}) or the octet ({\bf 8}) representation, 
the 4Q state $\rm |T\rangle$ can be rewritten as 
\begin{eqnarray}
| {\rm T} \rangle 
&=& C_1 |({\rm Q_1\bar{Q}_3})_1({\rm Q_2\bar{Q}_4})_1 \rangle \nonumber \\
&+& C_8 |({\rm Q_1\bar{Q}_3})_8({\rm Q_2\bar{Q}_4})_8 \rangle_1 
\end{eqnarray}
with appropriate constants $C_1$ and $C_8$ satisfying 
\begin{eqnarray}
|C_1|^2+|C_8|^2=1.
\end{eqnarray} 
After some calculation, one finds
\begin{eqnarray}
|{\bf 11}\rangle &\equiv&
|({\rm Q_1\bar{Q}_3})_1({\rm Q_2\bar{Q}_4})_1 \rangle =
\frac{1}{3}|{\rm Q}_1^a{\rm Q}_2^b \bar{\rm Q}_3^b \bar{\rm Q}_4^b \rangle, 
\label{A-six}
\\
|{\bf 88}\rangle &\equiv&
|({\rm Q_1\bar{Q}_3})_8({\rm Q_2\bar{Q}_4})_8 \rangle_1 \nonumber \\
&=& \frac{1}{2\sqrt{2}}
\{|{\rm Q}_1^a {\rm Q}_2^b \bar{\rm Q}_3^b \bar{\rm Q}_4^a\rangle 
-\frac{1}{3}|{\rm Q}_1^a {\rm Q}_2^b \bar{\rm Q}_3^a \bar{\rm Q}_4^b \rangle \},~~~
\label{A-seven}
\end{eqnarray}
which satisfy the orthonormal condition,  
\begin{eqnarray}
\langle{\bf 11}|{\bf 11}\rangle=\langle{\bf 88}|{\bf 88}\rangle=1, \ \langle{\bf 11}|{\bf 88}\rangle=0. 
\end{eqnarray}
Using Eqs.(\ref{A-one}), (\ref{A-six}) and (\ref{A-seven}), 
$C_1$ and $C_8$ can be obtained as  
\begin{eqnarray}
C_1&=&\langle {\bf 11} |{\rm T} \rangle \nonumber \\
&=&\frac{1}{6\sqrt{3}}
\epsilon^{abc}\epsilon^{ab'c'}
\langle {\rm Q}_1^d{\rm Q}_2^e \bar{\rm Q}_3^d \bar{\rm Q}_4^e |
{\rm Q}_1^b {\rm Q}_2^c \bar{\rm Q}_3^{b'} \bar{\rm Q}_4^{c'}\rangle \nonumber \\
&=&\frac{1}{6\sqrt{3}}
\epsilon^{abc}\epsilon^{abc}=\frac{1}{\sqrt{3}}, 
\end{eqnarray}
\begin{eqnarray}
C_8&=&\langle {\bf 88} |{\rm T} \rangle \nonumber \\
&=&\frac{1}{4\sqrt{6}}
\epsilon^{abc}\epsilon^{ab'c'}
\langle
{\rm Q}_1^d {\rm Q}_2^e \bar{\rm Q}_3^e \bar{\rm Q}_4^d|
{\rm Q}_1^b {\rm Q}_2^c \bar{\rm Q}_3^{b'} \bar{\rm Q}_4^{c'}\rangle \nonumber \\
&-&\frac{1}{12\sqrt{6}}
\epsilon^{abc}\epsilon^{ab'c'}
\langle 
{\rm Q}_1^d {\rm Q}_2^e \bar{\rm Q}_3^d \bar{\rm Q}_4^e|
{\rm Q}_1^b {\rm Q}_2^c \bar{\rm Q}_3^{b'} \bar{\rm Q}_4^{c'}\rangle \nonumber \\
&=&\frac{1}{4\sqrt{6}}\epsilon^{abc}\epsilon^{acb}
-\frac{1}{12\sqrt{6}}\epsilon^{abc}\epsilon^{abc}
=-\sqrt{\frac23}.
\end{eqnarray}
Then, using Eq.(\ref{A-one}), $C_2({\bf 1})=0$ and $C_2({\bf 8})=3$, we get  
\begin{eqnarray}
\langle {\rm T} | T_1^a T_3^a | {\rm T} \rangle 
&=&   |C_1|^2 \langle {\bf 11} | T_1^a T_3^a | {\bf 11} \rangle
    + |C_8|^2 \langle {\bf 88} | T_1^a T_3^a | {\bf 88} \rangle \nonumber\\
&=& |C_1|^2 \{ \frac12 C_2({\bf 1})-\frac43 \} +|C_8|^2 \{ \frac12 C_2({\bf 8})-\frac43 \} \nonumber \\
&=& -\frac{1}{3}.
\end{eqnarray}

In this way, for the Coulomb interaction in the 4Q system as shown in Fig.3, we obtain 
\begin{eqnarray}
\langle {\rm T} | T_1^a T_2^a | {\rm T} \rangle&=&\langle {\rm T} | T_3^a T_4^a | {\rm T} \rangle= -\frac{2}{3}, \\
\langle {\rm T} | T_1^a T_3^a | {\rm T} \rangle&=&\langle {\rm T} | T_1^a T_4^a | {\rm T} \rangle
=\langle {\rm T} | T_2^a T_3^a | {\rm T} \rangle \nonumber \\
&=&\langle {\rm T} | T_2^a T_4^a | {\rm T} \rangle= -\frac{1}{3},
\end{eqnarray}
and derive the Coulomb terms in Eq.(\ref{Vc4Q}).

\end{document}